\documentclass[hyper,a4paper]{JHEP}
\usepackage{amsmath,graphicx,epsfig,amssymb,multirow}
\allowdisplaybreaks
\def\gsim{\raise0.3ex\hbox{$\;>$\kern-0.75em\raise-1.1ex\hbox{$\sim\;$}}}
\def\lsim{\raise0.3ex\hbox{$\;<$\kern-0.75em\raise-1.1ex\hbox{$\sim\;$}}}

\preprint{arXiv:1005.xxxx\\ \today}
\title{Spin Discrimination in Three-Body Decays}
\author{Lisa Edelh\"auser$^{1,a}$, Werner Porod$^{1,2,b}$, Ritesh K. Singh$^{1,c}$ \\
$^1$Institut f\"ur Theoretische Physik und Astrophysik, Universit\"at W\"urzburg,\\
D-97074  W\"urzburg, GERMANY\\[1mm]
$^2$AHEP Group, Instituto de F\'{\i}sica Corpuscular --
    C.S.I.C./Universitat de Val{\`e}ncia \\
  E--46071 Val{\`e}ncia, Spain\\[1mm]
Email:~$^a$\email{ledelhaeuser@physik.uni-wuerzburg.de},\\
$^b$\email{porod@physik.uni-wuerzburg.de},\\
$^c$\email{singh@physik.uni-wuerzburg.de} }

\abstract{The identification of the correct model for physics
beyond the Standard Model requires the determination of the spin of
new particles. We investigate to which extent the spin of a new
particle $X$  can be identified in scenarios where  it decays 
dominantly in three-body decays $X\rightarrow f\overline{f} Y$.
Here we assume that $Y$ is a candidate for dark matter and escapes
direct detection at a high energy collider such as the LHC. We show that in the 
case that all  intermediate particles are heavy,
one can get information on the spins of $X$ and $Y$ at the
LHC by exploiting the invariant mass distribution of the
two standard model fermions. We develop a
  model-independent strategy to determine the spins without prior knowledge
  of the unknown couplings and test it in a series of Monte Carlo studies. 
  }

\begin{document}

\section{Introduction}

With the start of the Large Hadron Collider (LHC) the direct exploration of physics at the TeV scale
has begun. The hunt for new physics is one of the major topics in the experimental
program of the LHC. Many of these models predict partners of the known standard model (SM) particles,
which usually have the same quantum numbers and properties but for the mass and the
spin assignment. For example, in supersymmetric (SUSY) models the fermions have scalar
partners whereas in models with universal extra dimensions (UED) fermionic partners
are predicted. Due to the astrophysical requirement of explaining the dark matter
(DM) relic density of the universe, these models usually invoke an additional discrete
symmetry leading to a new stable particle which in general escapes detection at
future collider experiments. Examples are R-parity in SUSY models or
Kaluza-Klein-parity in UED where e.g.~the lightest neutralino or the lightest
Kaluza-Klein (KK) excitation of the vector bosons is the corresponding DM candidate,
respectively. The generic signature at LHC are in both cases  SM-fermions with high
transverse momentum and missing energy stemming from the escaping DM candidate.

An important question is:  How can one distinguish between different models? 
 These  models differ in the spins of the predicted new particles and, thus,
one has to develop methods to get information on the spin. First attempts have
been made for $s$-channel resonances \cite{Choi:2002jk,Alves:2008up,Osland:2008sy,Osland:2009tn} and
in case of subsequent two-body decays of the new particles 
\cite{Barr:2005dz,Smillie:2005ar,Athanasiou:2006ef,Athanasiou:2006hv,Smillie:2006cd,%
Wang:2006hk,Meade:2006dw,Kilic:2007zk,Alves:2007xt,Rajaraman:2007ae,Cho:2008tj,%
Wang:2008sw,Burns:2008cp} where in many cases model
dependent assumptions had been made. An additional possibility to get information on the 
spin is cross section measurements provided one knows the representation of the particle
produced \cite{Kane:2008kw}, e.g.~whether it is a colour triplet or a colour octet.
Hardly any attempt has been made so far in case of three-body
decays but for the case of distinguishing a gluino from the
KK excitation of a gluon \cite{Csaki:2007xm} and the quantum interference
method \cite{Buckley:2007th,Buckley:2008pp,Buckley:2008eb,Boudjema:2009fz}
which, in principle, is also valid for three-body decays. The quantum interference method requires fully reconstructed events
which can be  achieved at the ILC and only in few processes at the LHC. In contrast,
our method in this paper does not require full reconstruction of events.
In this paper we start
a series of investigations on how one can extract information on
the spins of new particles in a model independent way
if three-body decays are dominating.

We concentrate here on the case of the direct production of a new particle
$X$ decaying via a three-body decay into two SM-fermions and a new invisible 
particle $Y$, which escapes detection. We will show that  the invariant mass
distribution of the two detectable fermions contains sufficient information to
extract the spins of the unknown particles in such a decay.

In this paper, we consider  cases where the
intermediate particles are very heavy compared to the decaying one.
Examples of such cases are e.g.~split SUSY with very heavy scalars 
\cite{Giudice:2004tc}, split UED \cite{Park:2009cs}  or 
higgsless supersymmetric models \cite{Knochel:2008ks}. As it turns out, 
in this limit it is possible to determine the spin of the decaying
particle and the invisible particle provided the masses of the new
particles are known to some extent.  In contrast to model dependent approaches
to spin determination \cite{Csaki:2007xm, Alves:2006df}, we follow a
bottom-up approach that can be used for all models with heavy intermediate
particles and only renormalizable operators.
The case of lighter intermediate particles will be presented in a subsequent
paper.

This paper is organised as follows:  We first present the basic ideas
using a toy model in section \ref{sec:approach}. Afterwards, we present
a convenient parametrisation of the differential width for the 
 three-body decays as the product of a phase space factor and a
 polynomial. 
In section \ref{sec:strategy} we  develop a strategy to determine
the spins exploiting specific properties of the coefficients of this polynomial. 
We also investigate the impact of  different decay topologies and  discuss qualitatively
the influence of the mass of the intermediate particle. In section
\ref{sec:montecarlo} we test our strategy with the help
of   Monte Carlo examples and conclude in section \ref{sec:conclusions}.
The formulas of the various coefficients are given in the appendix.


\section{Basic Idea and General Setting}
\label{sec:general}

\label{sec:approach}
We investigate decays of the type $X\rightarrow f\overline{f} Y$ where $X$ 
and $Y$ are new particles being either scalars, vector bosons or fermions.
Here we assume that all 2-body decays of $X$ are either kinematically
forbidden or at least loop-suppressed compared to the tree-level
three-body decays considered. As mentioned above we assume that
all off-shell particles, which we denote collectively by $I$,
mediating these decays are much heavier than $X$, e.g.\ $m_I \gg m_X$.
In practice it is sufficient to assume $m_I \gsim 5  m_X$ as we will show below.
In addition we assume that $Y$ is a colour singlet as it should
serve as potential dark matter candidate. 

We will be as general as possible by taking the most generic Lagrangian
with arbitrary couplings of ${\cal O}(1)$ and dimension 4 operators.
From this we calculate the widths for the decays of $Y$ assigning different
spins to $X$ and $Y$, respectively. To simplify the notation we will abbreviate the decays $S\rightarrow f \bar{f} S$, $S\rightarrow f \bar{f} V$,
$V\rightarrow f \bar{f} S$, $V\rightarrow f \bar{f} V$ and $F\rightarrow f \bar{f} F$
by $(S,S)$, $(S,V)$, $(V,S)$, $(V,V)$ and $(F,F)$, where $S$, $V$ and $F$ stand for
scalar, vector boson and fermion, respectively. Note, that
the fermionic case covers both, Dirac- and Majorana-fermions.

 After integrating over
the momentum of the escaping particle $Y$, we expand the differential widths
in powers of $\epsilon = m_X/m_I$ and give the resulting
expressions as a phase space factor times a
Laurent series (actually polynomials in most cases) of a 
dimensionless quantity $\hat s$
which is derived from the invariant mass 
 $s=(p_f + p_{\bar{f}})^2$,
$p_f$ and $p_{\bar{f}}$ are the four momenta of the
SM-fermions. Note, that $s = m^2_{f\bar{f}}$ in the case of subsequent
two body decays studied in the literature
\cite{Barr:2005dz,Smillie:2005ar,Athanasiou:2006ef,Athanasiou:2006hv%
,Smillie:2006cd,Wang:2008sw,Burns:2008cp,Allanach:2000kt}.

Most of the features can be understood by considering the decays of
a particle $X$ charged under $SU(3)$ into two massless quarks and $Y$ . Considering
coloured particles in the first place is motivated by the fact
that they in general have sizable cross sections at the LHC. Moreover,
due to gauge invariance only a subset of all topologies are allowed
which simplifies the obtained expressions considerably.
Therefore we will first discuss these cases. The additional features of 
either taking $X$ as an $SU(3)$ singlet and/or the case
that the SM-fermions being massive (i.e. top quarks) will be discussed 
afterwards.

The coefficients in the Laurent series depend obviously on
the couplings and masses of the particles involved and one
might ask if and  how one gets information on the spins without
knowing these quantities. It turns out we have to
assume in our approach that the mass of $Y$  and the mass difference 
$(m_X - m_Y)$   are known  within a given
 uncertainty but
in general we do not need any information on the underlying
couplings. The basic idea is that different spin
assignments lead to different relations between these
coefficients which can be exploited. There is however one
obstacle: one cannot exclude on logical grounds that there
is a 'conspiracy' between the couplings suppressing 
the dominant terms in the $\epsilon$ expansion. This complicates
life somewhat but even in that case relations between the coefficients are
maintained as discussed below.

\subsection{Dependence of the invariant mass distribution  on the
spins of unknown particles}

\label{subsec:dependence}

\begin{figure}[tp]
\centering
\begin{picture}(200,150)
\put(0,0){\includegraphics[width=0.5\textwidth]{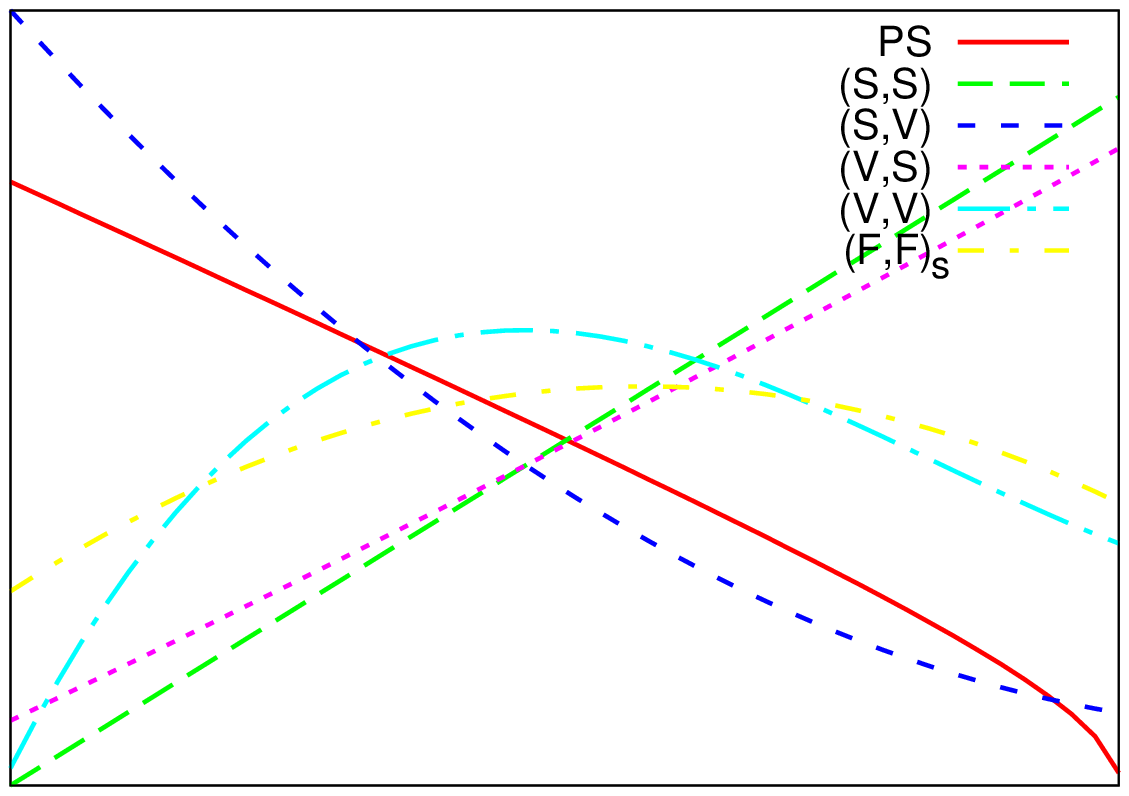}}
\put(-45,70){$\frac{d\Gamma}{ds}:PS$}
\put(110,-5){$s$}
\put(10,-5){$0$}
\put(200,-5){$s_{max}$}
\end{picture}
\caption{Differential width divided by a phase space factor $PS$ 
for the different decays 
$X\rightarrow f\overline{f} Y$, $X,Y \in S,V$ taking $m_f=0$, $m_X/m_Y=0.1$
and all couplings equal. In addition the phase space factor is drawn.
}
\label{fig:massdist} 
\end{figure}

We first discuss a set of toy models where the new particles $X$ and $Y$
are either scalars and/or vector bosons coupling to massless $u$-quarks and an additional
heavy fermion which we assume to be a Majorana-fermion. The invariant fermion
mass squared is $s=(p_f+p_{\bar{f}})^2 =E_1 E_2 \cdot (1-\cos{\theta_{f\bar{f}}})$ where
$E_i$ are the SM fermion energies and $\theta_{f\bar{f}}$ is the angle between them
in the rest-frame of $X$.
In Figure~\ref{fig:massdist} we show the differential decay width divided by
the phase space factor which is shown independently (red/full line). The behaviour
of the different curves can be understood using helicity and spin arguments.

The SM-fermions have definite helicity states as we have assumed them to be
massless. There are two kinematical configurations where the spatial angular momentum
in the rest-frame of $X$ 
is zero corresponding to $\cos{\theta_{f\bar{f}}} = \pm 1$: (i) The particle
$Y$ is at rest and the two fermions are back to back corresponding to
$\cos{\theta_{f\bar{f}}} = - 1$ with $s=s_{max}$. 
In this case the total spin of the fermions
sums up to one if it is either an $u_L \bar{u}_L$ or $u_R \bar{u}_R$ 
combination whereas
the total spin of the fermions is zero for the $u_L \bar{u}_R$ and $u_R \bar{u}_L$
combinations. Here we have introduced $u_{L,R} = P_{L,R}~ u$ with $P_{L,R} =
(1 \mp \gamma_5)/2$. (ii) The particle $Y$ flies opposite to the two fermions which
are flying parallel now corresponding to
$\cos{\theta_{f\bar{f}}} =  1$ with $s_{min}=0$.
 In this case the total spin of the fermions
sums up to one if it is either a $u_L \bar{u}_R$ or a $u_R \bar{u}_L$
combination whereas
the total spin of the fermions is zero for the  $u_L \bar{u}_L$ and $u_R \bar{u}_R$ 
combinations.

\paragraph{$(S,S)$: \\}

The matrix element $M_{fi}$ for this decay has
the generic form 
\begin{eqnarray*}
M_{fi}  =  \overline{u}_u\,\, (g_r P_R +g_l P_L) S_I (n^*_l P_R+n_r^* P_L) v_u 
\simeq  \frac{1}{m_I} \overline{u}_u (g_r n_l^* P_R + g_l n_r^* P_L)  v_u
\end{eqnarray*}
corresponding to the $u_L \bar{u}_R$ and $u_R \bar{u}_L$ combinations
since in the limit of a very heavy intermediate particle the corresponding propagator reduces to
 $S_I = 1/(p\!\!\!/ - m_I) \simeq 1/m_I$. As the total angular momentum of the final
state has to be zero, the matrix element has to vanish in case 
$\cos{\theta_{f\bar{f}}} =  1$ ($s=0$) as can also be seen in Fig.~\ref{fig:massdist}
whereas
for $\cos{\theta_{f\bar{f}}} =  -1$ the helicity assignment yields a non-vanishing 
matrix element. In the plot we have taken for all cases $n_l=n_r=g_l=g_r$.

\paragraph{$(S,V)$:\\}
This process has  a more involved structure
since the vector boson can have polarisation $\pm 1,0$. 
Using the matrix element one sees that
in contrast to the previous case one expects a non-vanishing matrix
element  for all $s$ 
\begin{eqnarray*}
M_{fi} & = & \overline{u}_u\,\, (g_r P_R +g_l P_L) S_I \gamma_\mu
(n^*_l P_R+n_r^* P_L) v_u \epsilon_Y^\mu \nonumber\\
&\simeq & \frac{1}{m_I} \overline{u}_u \gamma_\mu (g_r n_r^* P_L + g_l n_l^* P_R)  v_u
\epsilon_Y^\mu.
\end{eqnarray*}
These are the $u_L \bar{u}_L$ and $u_R \bar{u}_R$  combinations where for $s=0$
the total spin of the fermions is zero and for $s=s_{max}$
it is one.  In the limit $\theta_{f\overline{f}}\rightarrow 0$ ($s\rightarrow 0$) the amplitude
is proportional to $((m_X/m_Y)^2-1)$ ($m_X^2-m_Y^2$). The first factor is
larger and diverges as 
$m_Y\rightarrow 0$ which reflects the longitudinal component of the vector boson. 
This can be nicely seen in the blue (dashed) line
in Fig.~\ref{fig:massdist}.

\paragraph{$(V,S)$:\\}
In this case, we start with a spin 1 boson. A similar reasoning
as before  shows that the general trend should be opposite to
the $S\rightarrow f\bar{f}V$ case which is confirmed by Fig.~\ref{fig:massdist}
(pink/small dashed line).

\paragraph{$(V,V)$:\\}
This decay can have several helicity combinations which we want to sketch here. 
The matrix reads 
\begin{eqnarray*}
M_{fi} & = & \overline{u}_u\,\,\gamma_\mu (g_r P_R +g_l P_L) S_I
\gamma_\nu (n^*_r P_L+n_l^* P_R) v_u \epsilon_Y^\nu \epsilon_X^{\mu*} \nonumber \\
& \simeq & \frac{1}{m_I}
 \overline{u}_u\,\,\gamma_\mu \gamma_\nu (g_r n^*_r P_L + g_l n^*_l P_l)  v_u
  \epsilon_Y^\nu \epsilon_X^{\mu*}
  \label{eq:vffv}
\end{eqnarray*}
As in the $(S,S)$ case we find the 
$u_L \bar{u}_R$ and $u_R \bar{u}_L$ combinations. However, now in principle
for all $s$ one can expect a non-vanishing matrix element squared. However,
as can be seen in Fig.~\ref{fig:massdist} it can be zero for $s=0$ for
special combinations of the couplings, e.g.~as in our case with
$n_l=n_r=g_l=g_r$ (cyan/dot-dashed line). 
In this sense the example shown is an extreme case and in general the matrix
element will be non-zero for $s=0$.

\subsection{General structure of the differential widths}
\label{sec:expansion}

We now discuss  the general structure for the  decays 
\begin{eqnarray}
X (p,m_X)&\rightarrow f(q_1,m_f) \,+ \overline{f}(q_2,m_f)\, + \, Y(q_3,m_Y)
\label{eq:simpledecay}
\end{eqnarray}
with $X$ and $Y$ being scalars $(S)$, vectors $(V)$
 or fermions $(F)$ 
and  heavy intermediate particles $I$ with mass $m_I$. 
We will consider several contributions at the same time and assume that
the masses of all intermediate particles are equal to maximise interference
effects which usually complicate things.
Beside  the usual Mandelstam variables
\begin{eqnarray*}
& &  s = (p-q_3)^2=(q_1+q_2)^2\,\,;\,\,
t = (p-q_2)^2=(q_1+q_3)^2\\
\mbox{and} &&  u = (p-q_1)^2=(q_2+q_3)^2 = -s-t-m_{X}^2-m_{Y}^2+2 m_f^2
\end{eqnarray*}
we introduce the dimensionless parameters $\tau_i$ and $\hat{s}$
\begin{equation}
\hat{s}= \frac{ \left(4 \tau_f^2+(\tau_Y-1)^2\right)- \frac{2 s}{m_X^2}}
              {\left(4 \tau_f^2-(\tau_Y-1)^2\right)} 
\,\,;\,\,
\tau_Y  =  \frac{m_Y}{m_X}
\,\,;\,\,
\tau_f  = \frac{m_f}{m_X}
\,\,;\,\,
\tau_C=\frac{M_C}{m_X}
\label{eq:shortcuts}
\end{equation}
where $M_C$ denotes dimensionful couplings, e.g.~as they appear in
the $ZZH$ vertex or the trilinear soft SUSY breaking parameters $A_i$.
We note that $\hat{s}_{min}=-1$ and $\hat{s}_{max}=1$.

We expand the matrix elements squared in powers of 
$\epsilon = m_X/m_I$ and perform an integration over $t$
as in this way we integrate over the momentum of the unobserved particle $Y$.
For the $t$-integration we find the boundaries
\begin{eqnarray}
t_\pm&=& \frac{1}{4} m_X^2 \Bigg((\tau_Y+1)^2+\hat{s} \left(4
   \tau_f^2-(\tau_Y-1)^2\right)
\nonumber\\
&\pm& \left((\tau_Y-1)^2-4
   \tau_f^2\right)\times \left.
\sqrt{\frac{(1-\hat{s}^2)
   \left((1-\hat{s})( (\tau_Y-1)^2- 4 \tau_f^2)+ 8\tau_Y \right)}{4
   \tau_f^2+(\tau_Y-1)^2+\hat{s} \left((\tau_Y-1)^2-4 \tau_f^2\right)}}\right)\\
PS & = &\int_{t_-}^{t_+}  d t \nonumber \\
& =& \left((\tau_Y-1)^2-4
   \tau_f^2\right)
\frac{m_X^2}{2} \sqrt{\frac{ (1-\hat{s}^2) 
   \left((1-\hat{s})((\tau_Y-1)^2- 4 \tau_f^2)+8 \tau_Y \right)
   }{4
   \tau_f^2+(\tau_Y-1)^2+\hat{s} \left((\tau_Y-1)^2-4\tau_f^2\right)}}
\label{eq:phasespace}
\end{eqnarray}
where we have also defined the 'phase space' PS.
In this way the differential decay rate can be written as
\begin{eqnarray}
\frac{d\Gamma}{d \hat{s}} & = & 
\frac{PS}{(2\pi)^3\,\, 256 \,\,m_X} 
\left(\frac{Z}{(a\hat{s}+b)^2}+\frac{A}{a\hat{s}+b}+ B + C\cdot \hat{s}
+D\cdot \hat{s}^2     + E \cdot\hat{s}^3 + F \cdot\hat{s}^4\right)\nonumber\\
\label{eq:decayrate2}
\end{eqnarray}
where $a=\left((\tau_Y-1)^2-4\tau_f^2\right)$ and 
$b=\left((\tau_Y-1)^2+4\tau_f^2\right)$.
 The prefactors $Z,A,\ldots,F$ are functions of $\epsilon$, the
$\tau_i$ and the couplings.  Note that $Z$ and $F$ only appear
in the case of the decay $(V,V)$.
As we exemplify the main features for massless SM-fermions, we display Eq.~(\ref{eq:phasespace}) 
 for this case:
\begin{equation}
\begin{array}{l}
t_\pm =
\frac{1}{4} m_X^2 \left((\tau_Y+1)^2-\hat{s} (\tau_Y-1)^2\pm(1-\tau_Y)
\sqrt{(1-\hat{s}) \left((1-\hat{s}) (\tau_Y-1)^2 + 8\tau_Y\right)} \right)
\\[2ex]
\mbox{PS} =
\frac{1}{2} m_X^2 (1-\tau_Y) \sqrt{(1-\hat{s}) 
\left((1-\hat{s}) (\tau_Y-1)^2 + 8\tau_Y \right)
}
\end{array}
 \label{eq:phasespace2}
\end{equation}

\subsubsection{Decays of bosons}

\label{sec:setup}
As one can see from Table \ref{tab:alltops}, there are three 'topologies'
which contribute differently to the decay rate. The second 'topology' only
contributes if both, the $Y$- and the $X$-particle, are their own anti-particles.
Obviously, topologies 1 and 2 will in general contribute at $\mathcal{O}(\epsilon^2)$ 
whereas the third one in general only at $\mathcal{O}(\epsilon^4)$ due to the different structures
of the propagators. Only in the case where the dimensionfull scalar-vector-vector  or 
triple scalar couplings are of $\mathcal{O}(m_I)$, the third topology might 
contribute at a smaller power of $\epsilon$
as will be discussed below.
In the further calculation we neglect terms
higher than $\mathcal{O}(\epsilon^4)$.
\begin{table}
\begin{tabular}{l|r|r|rr}
\hline
Decay & Top. 1 & Top. 2 & \multicolumn{2}{c}{Top. 3(s/v) } \\ 
\hline
\raisebox{7ex}
{
$(S,S)$
}
 &
\includegraphics[height=0.2\textwidth]{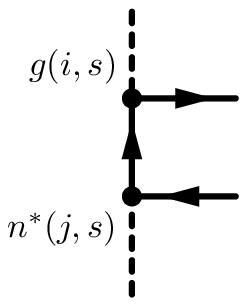}&
\includegraphics[height=0.2\textwidth]{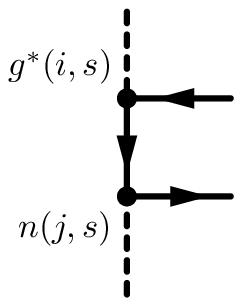}&
\includegraphics[height=0.2\textwidth]{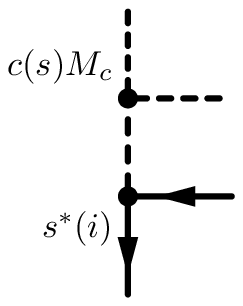}&
\includegraphics[height=0.2\textwidth]{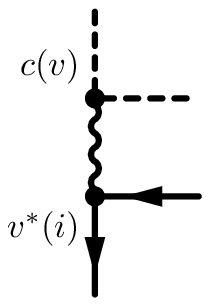}
\\ \hline
\raisebox{7ex}
{
$(S,V)$
}
&\includegraphics[height=0.2\textwidth]{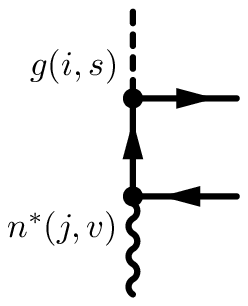}&
\includegraphics[height=0.2\textwidth]{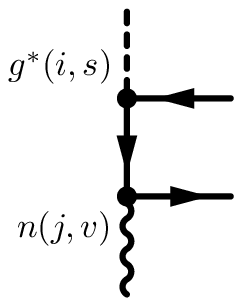}&
\includegraphics[height=0.2\textwidth]{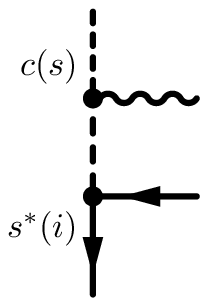}&
\includegraphics[height=0.2\textwidth]{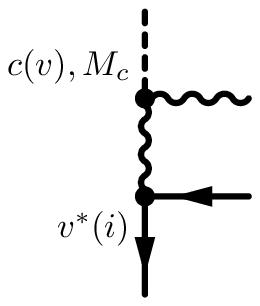}
\\ \hline
\raisebox{7ex}{
$(V,S)$
} &
\includegraphics[height=0.2\textwidth]{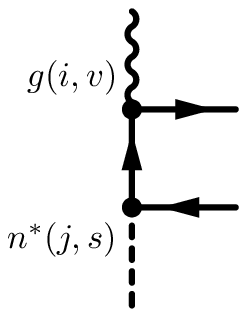}&
\includegraphics[height=0.2\textwidth]{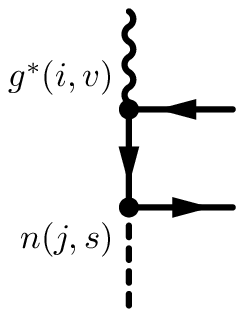}&
\includegraphics[height=0.2\textwidth]{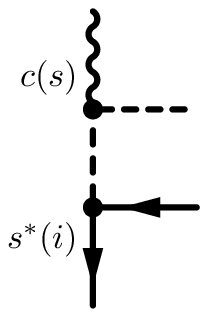}&
\includegraphics[height=0.2\textwidth]{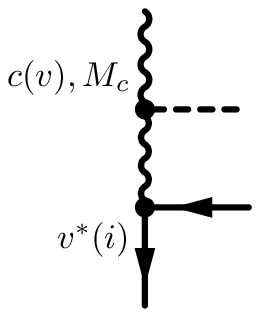}
\\ \hline
\raisebox{7ex}
{
$(V,V)$ }
&
\includegraphics[height=0.2\textwidth]{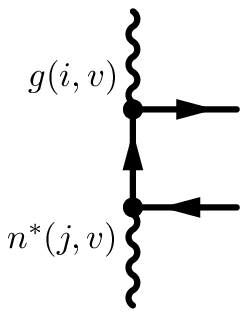}&
\includegraphics[height=0.2\textwidth]{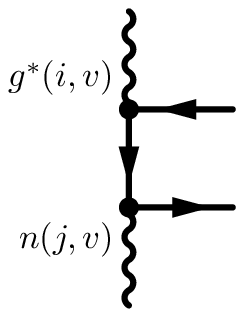}&
\includegraphics[height=0.2\textwidth]{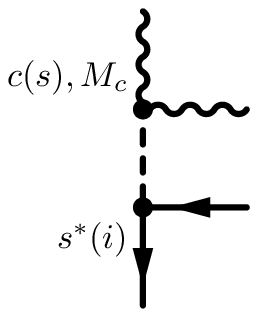}&
\includegraphics[height=0.2\textwidth]{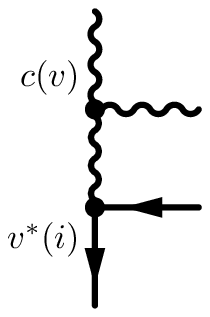}
\\ 
\end{tabular}

\begin{tabular}{c|c|c|c|c|c }
\hline
 Top.~1~(s) & Top.~2~(s) & Top.~3~(s) & Top.~1~(v) & Top.~2~(v) & Top.~3~(v) \\ 
\hline
\includegraphics[height=0.17\textwidth]{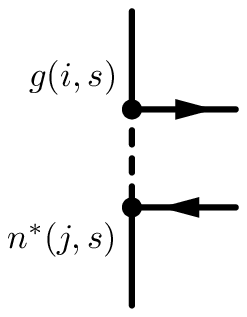}&
\includegraphics[height=0.17\textwidth]{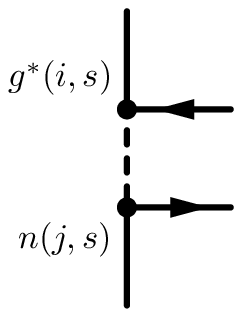}&
\includegraphics[height=0.17\textwidth]{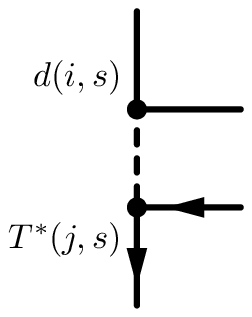}&
\includegraphics[height=0.17\textwidth]{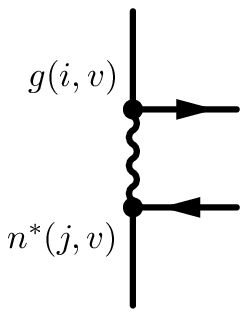}&
\includegraphics[height=0.17\textwidth]{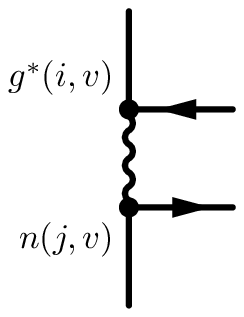}&
\includegraphics[height=0.17\textwidth]{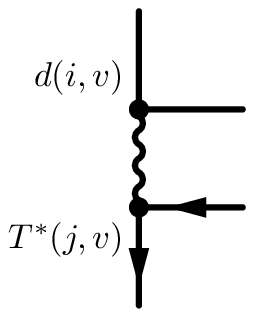}
\\ \hline
\end{tabular}
\caption{Topologies for the decays of bosons $X\rightarrow f \overline{f} Y$ (top) and
fermions  (F,F) (bottom) with $i,j \in \{l,r\}$
(see also Eq.~(\ref{eq:couplings1})).}
\label{tab:alltops}
 \end{table}

The generic Lagrangian density for these decays reads as
\begin{eqnarray}
\mathcal{L}_{i,j,k} & = & X_i   \overline{I}_f \widetilde{G}_i f
 + Y_i  \overline{I}_f \widetilde{N}_i f 
 + I_i \overline{f}  \widetilde{T}_i f 
 + I_i X_j Y_k\widetilde{\Gamma}_{ijk}+h.c.
\label{eq:genericxffy}
\end{eqnarray}
where $i,j,k=s,v$ and $I_i$ is the intermediate off-shell particle. The generic couplings
are given by 
\begin{eqnarray}
\begin{array}{lll}
 \widetilde{G}_i: & G_s = \left( g(r,s) P_R + g(l,s)P_L\right);
                  &G_v= \gamma^\mu\left( g(r,v) P_R + g(l,v)P_L\right)\\
\widetilde{N}_i: &  N_s= \left( n(r,s) P_R + n(l,s)P_L\right); 
                 &N_v= \gamma^\mu\left( n(r,v) P_R + n(l,v)P_L\right)\\
 \widetilde{T}_i: &    T_s= \left( s(r) P_R + s(l) P_L\right);
                  &T_v= \gamma^\mu\left( v(r) P_R + v(l)P_L\right)\\
\end{array}
\label{eq:couplings1}
\end{eqnarray}
and for $\widetilde{\Gamma}_{ijk}$
\begin{equation}
\begin{array}{ll}
    \Gamma_{sss} = c(s) M_C;& \Gamma_{vvs}=   \Gamma_{vsv}=\Gamma (v^\mu,v^\nu,s) = c(v) M_C g^{\mu\nu}\,\,\,\,\,\,\,\,\,\,\,\,\,\\ 
 \Gamma_{svv} =\Gamma (s,v^\mu,v^\nu)= c(s) M_C g^{\mu\nu};
 & \Gamma_{vs_1s_2} =\Gamma (v^\mu,s_1,s_2)= c(v) (p_{s_2}-p_{s_1})^\mu  \\
   \lefteqn{\Gamma_{s_1 s_2v}=  \Gamma_{s_1 v s_2}= \Gamma (s_1,s_2,v^\mu) = c(s)  (p_{s_2}-p_{s_1})^\mu }\\
 \Gamma_{v_1v_2v_3} = \Gamma (v_1^\nu,v_2^\rho,v_3^\mu)= \lefteqn{
c(v) ( (p_{v_1}-p_{v_2})^{\mu} g^{\nu\rho} + (p_{v_2}-p_{v_3})^{\nu} g^{\mu\rho} + (p_{v_3} -p_{v_1})^\rho g^{\mu\nu})} &\,\,\,\,\,
\end{array} 
\label{eq:couplings2}
\end{equation}
where the indices of the vertex expressions $\Gamma_{ijk} $ correspond to those of the Lagrangian in Eq.~\ref{eq:couplings1}.

 The matrix element including all topologies
 shown in Table \ref{tab:alltops} reads 
\begin{equation}
\begin{array}{lll}
\displaystyle\mathcal{M}_{i,j} & = & \,\,\,\, \overline{u}(q_1,m_f) G_i F_p\!\!\!\!\!/ \,\,\,\,\, \gamma^0 N_j^{\dagger}\gamma^0 \,\, v(q_2,m_f) \epsilon_{i}(p)\epsilon_{j}(q_3)\\[2ex]
 && \displaystyle+\overline{u}(q_1,m_f) N_j  F_p\!\!\!\!\!/ \,\,\,\,\, \gamma^0 G_i^{\dagger}\gamma^0\,\, v(q_2,m_f) \epsilon_{i}(p)\epsilon_{j}(q_3) \\[2ex]
 && \displaystyle+   \sum_{k=s,v}\Gamma_{kij} W_{P,i} \overline{u}(q_1,m_f)\gamma^0 T_k^\dagger\gamma^0 v(q_2,m_f)
\end{array}
\end{equation}
with the 'polarisation' vectors 
\begin{eqnarray*}
\epsilon_s  =  1\!\!1;\,\,\,&&\epsilon_v  =  \widetilde{\epsilon}^\mu
\end{eqnarray*} 
and the fermion and boson propagators
\begin{equation}
\begin{array}{ccc}
F_p\!\!\!\!\!\ =i\frac{p\!\!\!/ \,\,\,+m_I}{p^2-m_I^2};&\,\,\,\,\,
W_{P,s} =i\frac{1}{p^2-m_I^2}; &\,\,\,\,\, 
W_{P,v} =-i\frac{\left(g^{\mu\nu}-p^\mu p^\nu /m_I^2\right)}{p^2-m_I^2} 
\end{array}
\label{eq:propagators}
\end{equation}

\subsubsection{Decays of fermions}

In this case the  generic Lagrangian is given by 
\begin{eqnarray}
\mathcal{L}_i &=& I_i  \overline{f} G_i M_x
 + I_i  \overline{f} N_i M_y
 +  I_i  \overline{M_y} \Gamma_i M_x 
 + I_i \overline{f} T_i f+h.c.
\label{eq:genericmajo}
\end{eqnarray}
where $M_{x,y}$ denote the spinors of the new fermions and $i=s,v$ denotes whether
the exchanged particle is a scalar or a vector boson.
The couplings are similar to those above:
\begin{eqnarray}
\begin{array}{lll}
\mathcal{M}_i & = & \,\,\,\,\, \Big[ \overline{u}(q_1,m_f) G_i u(p,m_X)\Big] \,\,W_{p,i}\,\, \left[\overline{u}(q_3,m_Y) \gamma^0 N^\dagger_i \gamma^0 v(q_2,m_f) \right] \\[2ex]
&& + \Big[ \overline{u}(q_1,m_{f})N_i v(q_3,m_Y)\Big] \,\,W_{p,i}\,\, \left[ \overline{v}(p,m_{X})\gamma^0 G^\dagger_i \gamma^0 v(q_2,m_f)\right] \\[2ex]
 && +  \Big[\overline{u}(q_3,m_Y) \Gamma_{i} u(p,m_X)\Big] \,\, W_{p,i}\,\,\left[ \overline{u}(q_1,m_f) \gamma^0 T_i^\dagger \gamma^0 v(q_2,m_f)\right]
\end{array}
\end{eqnarray}
corresponding to the topologies given in Table~\ref{tab:alltops}
with the same couplings as in Eq. (\ref{eq:couplings1}, \ref{eq:couplings2}) and additionally:
\begin{eqnarray}
\begin{array}{lll}
i=s: && \Gamma_s =\left( d(s,r)P_R+ d(s,l)P_L\right)\\
i=v: && \Gamma_v =\gamma^\mu\left(d(v,r)P_R+ d(v,l)P_L\right)\\
\end{array}
\end{eqnarray}


\section{Strategy for Spin Identification}
\label{sec:strategy}
In this section we discuss the strategy for discriminating the various
scenarios with different spins assigned to the particles $X$ and $Y$.  The
procedure is to find suitable relations or to mark the signs of the
different coefficients $Z,A,\ldots,F$. This is done in the second
part of this chapter. Before this we will  have a
look at the different topologies and their contribution depending on
the chosen colour structure.

Our main focus here is on the case of massless fermions as this is already
sufficient to get the required information. This immediately implies
a considerable simplification because some of the coefficients for
the differential width are zero and we obtain
\begin{eqnarray}
\frac{d\Gamma}{d\hat{s}} =  \frac{PS}{(2\pi)^3\,\,256\,\,m_X}  \big(  B
+ C \hat{s} + D \hat{s}^2 + E \hat{s}^3 +F \hat{s}^4\big)
\label{eq:decayrate}
\end{eqnarray}
The formulas for the coefficients are given in the appendix
 for the corresponding lowest order
in $\epsilon$. For the case $(S,S)$  we also
give the higher orders up to $\epsilon^4$. 
Moreover, it
turns out that the  decays in top-quarks, the only SM fermion with mass
of $\mathcal{O}(100\mbox{ GeV}) $, behave in the same way and the
discrimination is also possible in this case as will be discussed at the end of
this section. This implies that one has a second system to test the spin assignments
in an independent way.

We start with a subset of the topologies given in Table~\ref{tab:alltops}, namely
topologies 1 and 2 as these are typically realised in extensions of
the SM, e.g.~in models with extra dimensions or in SUSY. Moreover,
we will first further restrict ourselves to  scenarios where
$X$ is charged under $SU(3)$, e.g.~a colour octet gluino or a 
KK excitation of a gluon, and $Y$ is electrically neutral and 
uncharged under $SU(3)$. This is motivated by the fact that the LHC is
a hadron collider. 
{In the second step we add the third topology. But it turns out that
in the case of scalar contributions to $(S,S)$ and $(S,V)$
gauge invariance allows only two additional terms because the SM-fermions
are charged under $SU(3)$.} 
However, these contributions will
in general be of order $\epsilon^4$ due to the boson propagator except 
for the case where the trilinear scalar
coupling is of  order $m_I$ in which there might be contributions at order
$\mathcal{O}(\epsilon^2)$. In the third step we will also discuss the complications and
their potential solutions in case that $X$ is an $SU(3)$ singlet.

\subsection{Signs of the coefficients}
\label{sec:coefficients}

It turns out that some of the coefficients have a definite sign independent
of the couplings and masses involved. This important fact will be used
later to discriminate between the different spin assignments of $X$ and $Y$.
We collected the signs of the different coefficients for all decays of bosons in
Table~\ref{tab:masslessdecaycoeff} 
where we have expanded the coefficients in powers of $\epsilon$, e.g.
\begin{equation}
B = \sum_{k=2}^4 B_k \epsilon^k
\end{equation}
Some of the signs in Table~\ref{tab:masslessdecaycoeff} are obtained
analytically but several are gained
numerically by scanning and inserting random couplings in the range
$[-1,1]$.  There are some coefficients where
the sign cannot be determined without knowing the mass ratios or the
couplings which are marked by "$\pm$". Moreover, we have
put a 0 whenever the coefficient itself vanishes.
We give the signs for three cases, ordered
from the most general one to the most restricted one:
(i) $X$ is an $SU(3)$ singlet, where all topologies of Table~\ref{tab:alltops} 
contribute. The corresponding columns are denoted by $s$. (ii) $X$ is charged under
$SU(3)$ and all possible contributions allowed are taken into account and
the corresponding columns are denoted by $c$.
(iii) $X$ is charged under
$SU(3)$ and only topologies 1+2 contribute and, thus, the corresponding
columns are denoted by 1+2. 
In the subsequent sections these cases will be discussed in the reversed order
focussing on the terms of order $\epsilon^2$.
The $\epsilon^4$ order
is only of interest, if the leading order is zero, which is the case
for some special coupling arrangements discussed in section~\ref{sec:faking}.  

\begin{table}[t]
\centering
\begin{tabular}{|c| c cc | cc  | cc | cc | }
\hline
 & \multicolumn{3}{c|}{$(S,S)$} & \multicolumn{2}{c|}{$(S,V)$} & \multicolumn{2}{c|}{$(V,S)$} & \multicolumn{2}{c|}{$(V,V)$} \\ \hline
$\epsilon^2$ & $s$	 & $c$ 	& 1+2 & $s$ & $c$/1+2 & $s$ & $c$/1+2 & $s$ & $c$/1+2 \\ \hline 
$B_2$ 	     &	$+$ 	 &$+$	& $+$	     & $+$	& $+$    	&$+$&$+$         	& $\pm$ & $\pm$      \\
$C_2$  	     &  $+$	 &$+$ 	& $+$	     &$\pm$	&$\pm$     	& $+$&$+$       	& $\pm$ &   $\pm$     \\
$D_2$        &  $0$ 	 &$0$	& $0$	     &$+$ 	&$+$         	&$+$&$+$       		& $\pm$ & $\pm$       \\
$E_2$        & 	$0$ 	 &$0$	& $0$	     &$0$ 	&$0$       	& $0$&$0$       	& $+$    &  $+$      \\
\hline
$\epsilon^3$ &      &    &         &         &              &            &              &            &          \\ \hline 
$B_3$     & $\pm$ &$\pm$	&$0$	   &$\pm$      &$0$          &$\pm$        &$0$          		&$\pm$     	&$0$          \\
$C_3$    & $\pm$ &$\pm$&$0$	   &$\pm$      &$0$          &$\pm$        &$0$                  &$\pm$         		&$0$          \\
$D_3$     &   $0$ &$0$	 &$0$	   &$\pm$     &$0$          &$\pm$        &$0$          &$\pm$         	&$0$          \\
$E_3$   &   $0$  &$0$   &$0$	   &$0$      &$0$          &$0$              &$0$        &$\pm$    		&$0$          \\
\hline
$\epsilon^4$ & 	    &    &         &      &              &            &              &         	&          \\ \hline 
$B_4$     & $+$  &$+$	 &$+$	   &$+$  	&$+$          &$+$        &$+$         	 &$+$     &   $+$          \\
$C_4$   & $\pm$	&$+$	 &$+$	   &$\pm$  	&$\pm$          &$\pm$        &$\pm$         	 &$\pm$     &   $\pm$          \\
$D_4$     & $\pm$&$-$	 &$-$	   &$\pm$  	&$\pm$          &$\pm$        &$\pm$         	 &$\pm$     &   $\pm$          \\
$E_4$   &   $0$&$0$	 &$0$	   &$\pm$  	&$\pm$          &$\pm$        &$\pm$         	 &$\pm$      &  $\pm$             \\
$F_4$     &   $0$&$0$	 &$0$	   &$0$  	 &$0$          &$0$        &$0$          	&$\pm$      &  $\pm$             \\
\hline
\end{tabular}
\caption{
Signs of the coefficients for the case of a boson decaying into another 
boson and massless SM-fermions 
in the final state for different
powers of $\epsilon$. The rows correspond to the cases: ($s$) $X$ is an $SU(3)$ singlet,
($c$) $X$ is charged under $SU(3)$ taking all possible topologies into account 
and ($1+2$)  $X$ is charged under $SU(3)$ taking topologies 1+2 into account.
 The $\pm$ marks the cases 
where the sign cannot be determined without 
knowing the masses/couplings and $0$ marks the cases with a  vanishing coefficient.}
\label{tab:masslessdecaycoeff}
\end{table}

In case of a new fermion $X$ it turns out that the result does neither
depend on the spin of the exchanged particle nor on the topology, e.g.~it
does not matter if all topologies are taken or only a subset.
Since  we have only bosonic propagators 
there are  only the $\mathcal{O}(\epsilon^4)$ contributions and we find:
\begin{eqnarray}
\mathrm{sign}(B_4)=+ \hspace{3mm},\hspace{3mm}
\mathrm{sign}(C_4) = \pm \hspace{3mm},\hspace{3mm}
\mathrm{sign}(D_4) = - \hspace{3mm},\hspace{3mm}
\mathrm{sign}(E_4)=0
\label{eq:signMFFM}
\end{eqnarray}

\subsection{Decays into massless SM-fermions in case of topologies 1+2}
\label{subsec:masslessstrategy}

\begin{figure}[t]
\centering
\includegraphics[width=0.9\textwidth]{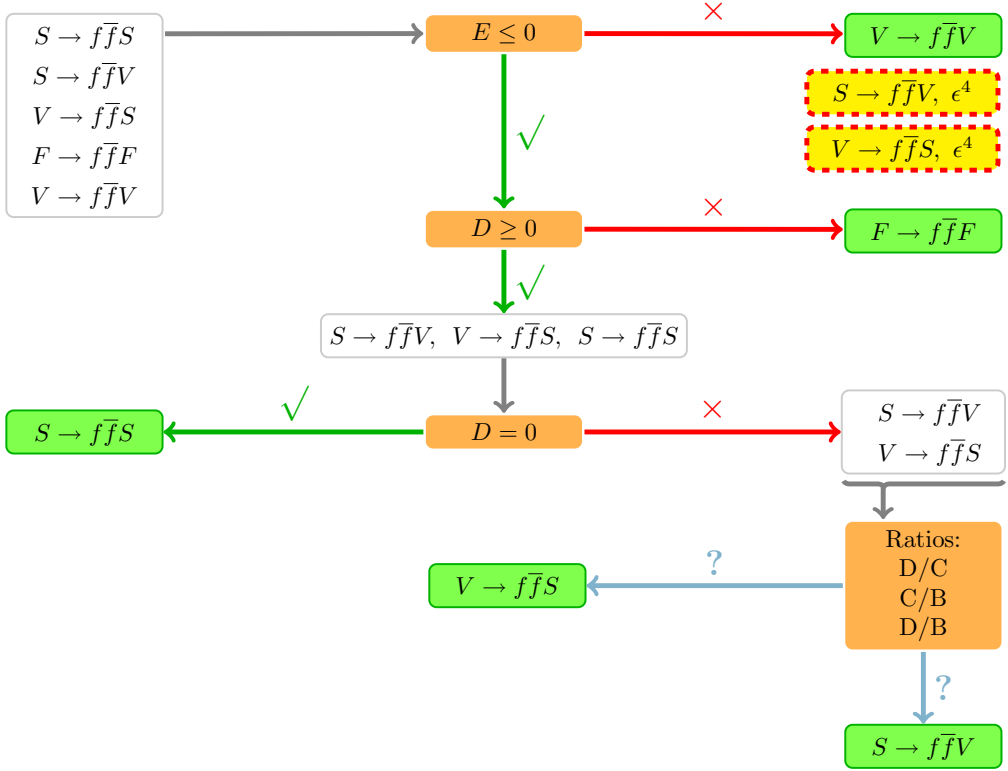}
\caption{Flowchart for the strategy to discriminate 
  different spin assignments for $m_f=0$ using the signs of
  the coefficients given in section~\ref{sec:coefficients}. 
  Green (solid frame) boxes are for the case of taking only
  the leading order into account. Impact or dominance of
  higher order terms is given by the yellow (dashed frame) boxes, see text
  for details.}
\label{fig:strategy}
\end{figure}

Let's assume that we have measured the differential decay width of a new particle and
determined the coefficients introduced above accurately in a fit. 
In section \ref{sec:montecarlo} we will discuss first Monte Carlo studies
at the parton level where we also review the obtainable accuracy.
This  can be combined with our knowledge
 on the various coefficients introduced so  far to 
determine the spins of the new particles or at least to exclude certain
possibilities. The main strategy is summarised in Fig.~\ref{fig:strategy}
and explained in some more detail below.  

Let's start with the $E$ term which is
only non-zero in the $(V,V)$ case. This immediately implies that $(V,V)$
is preferred once the 'measured' $E$ term is larger than 0. 
For consistency we check that $B>0$. The next
step is to look at the $D$ term as for $D\neq 0$ and $E=0$ the sign of
$D$ determines whether one is dealing with fermions $(D<0)$ or bosons
$(D \ge 0)$ where the latter case includes  $(S,S)$,  $(S,V)$ and $(V,S)$.
In case of $D=0$ only the case $(S,S)$ remains. 
To further distinguish the cases $(S,V)$ and $(V,S)$  from each
other, one has to consider the ratios:
\begin{equation}
\begin{array}{lll}
(S,V): & D/C= \frac{(\tau_Y-1)^2}{22 \tau_Y^2-4 \tau_Y-2} &\,\,\, \in
 [-\infty,-\frac{1}{3}] \cup [0,\infty] \\
(V,S): & D/C=-\frac{(\tau_Y-1)^2}{2 (\tau_Y (\tau_Y+2)-11)}  & \,\,\, \in [0,\frac{1}{22}] \\[2ex]
(S,V): & C/B=  \frac{22 \tau_Y^2-4 \tau_Y-2}{\tau_Y (25
   \tau_Y+6)+1}&\,\,\, \in [-2,\frac{1}{2}] \\
(V,S): & C/B= \frac{8 (\tau_Y+9)}{\tau_Y (\tau_Y+6)+25}-2 & \,\,\, \in[\frac{1}{2},\frac{22}{25}]\\[2ex]  
(S,V): & D/B= \frac{(\tau_Y-1)^2}{\tau_Y (25 \tau_Y+6)+1} & \,\,\, \in [0,1] \\
(V,S): & D/B= \frac{(\tau_Y-1)^2}{\tau_Y (\tau_Y+6)+25} & \,\,\, \in  [0,\frac{1}{25}]
\end{array}
\label{eq:ratiosmassless}
\end{equation}
since here the dependence on the unknown couplings cancel as can be seen
from eqs.~(\ref{eq:sv}) and
(\ref{eq:vs}). For these
decays we have three possible ratios shown in Eq.~(\ref{eq:ratiosmassless}).
We see that the ratio $C/B$ has no overlap and hence best suited ratio to 
distinguish between $(S,V)$ and $(V,S)$.
The $D/B$ and $D/C$ ratios seem to be less useful since the intervals overlap,
but except for $\tau_Y=1$, e.g.~$m_Y=m_X$, but they are never equal for 
$\tau_Y \neq 1$.  In the range
where $\tau_Y$ is close to one, the SM-fermions become very soft and this
part will be excluded because a lower cut on their energies is
put in practice. 
 Last but not least we note
that the $(S,S)$ can be further checked by the requirement that $B/C=1$ as
can be seen from Eq.~(\ref{eq:ss}).
Therefore, independent of the mass ratios, one can state that all those five
cases can be discriminated from each other. 

\subsection{Impact of the third topology}
\label{sec:third}

We have seen  in section~\ref{sec:expansion}
that the dominant contributions stem from topologies 1 and 2  of Table~\ref{tab:alltops}
in case of decays of bosons because the third topology generally
contributes at $\mathcal{O}(\epsilon^4)$. One might asked if one
of the dimensionful couplings in the diagrams of the
third topology can become so large to disturb
the above strategy. Note, that in case of a new fermions higher orders
in $\epsilon$ have no impact and, thus, we restrict the discussion here to 
 decays of bosons.

 Let us first consider the case
that $X$ is in a non-trivial $SU(3)$ representation.
Here only in the $(S,S)$ there is a potentially dangerous contribution
because the trilinear scalar coupling can in principle be of order
$m_I$. However, from Eq.~(\ref{eq:ss}) we see that only the equality
$B=C$ gets broken, the $D$-term will only get a tiny contribution, and thus
the general strategy should still work. In case of the scalar exchange in
the third topology of the $(S,V)$ case the coupling is momentum dependent
and is of order $m_X$ and one is safe again. 
In the case $X$ is a colour singlet all diagrams for the third
topology in Table~\ref{tab:alltops} contribute in principle. 
Here one has to distinguish two cases:
(i) there are no new vector bosons or the new vector bosons do not
belong to a new gauge group. In this case a detailed inspection  
of the diagrams shows that one arrives at the same conclusions as above, because
all  dimensionful couplings, which had not been considered before,
 have to be of the order $m_X$
due to $SU_L(2)$ gauge invariance. 
(ii) There is a new gauge group at higher energies to which the intermediate
vector bosons belong. In this case the SVV coupling as well
as the masses of the vector boson will be of same order of magnitude and, thus,
our assumption that the intermediate particle is much heavier than the decaying
one does not hold in this case.

\subsection{Special combinations of couplings}
\label{sec:faking}

Up to now we have considered only the leading terms in case of decays
of new bosons. However, it can happen
that for special helicity
assignments, the leading order $\epsilon^2$ becomes zero. An example
is the $(S,S)$ case as can be seen  in eqs.~(\ref{eq:ss}) and (\ref{eq:ss3})
where the leading order is proportional to $(g(r,s) n(l,s) +  g(l,s) n(r,s))^2$ which is 
0 in the case of e.g. $g(l,s)=n(r,s)=0$. 
In such cases the  $p\!\!/ /m^2_I$ part of the fermion propagator  becomes important 
which contributes only at $\mathcal{O}(\epsilon^4)$. The question now is to which
extent we would arrive at wrong conclusions using the strategy discussed so far.
In table~\ref{tab:eps4} we give the resulting signs of the coefficients which have 
to be compared with the $\mathcal{O}(\epsilon^2)$ coefficients of
 Table~\ref{tab:masslessdecaycoeff}.

\begin{table}[t]
\centering
\begin{tabular}{|c| c cc | cc  | cc | cc | }
\hline
 & \multicolumn{3}{c|}{$(S,S)$} & \multicolumn{2}{c|}{$(S,V)$} & \multicolumn{2}{c|}{$(V,S)$} & \multicolumn{2}{c|}{$(V,V)$} \\ \hline
$\epsilon^4$ 	& $s$ 	& $c$	& 1+2 	& $s$ 	& $c$/ 1+2 & $s$ & $c$/1+2 & $s$ & $c$/1+2 \\ \hline 
$B_4$ 		&$+$	&$+$	&$0$	&$+$	&$+$	&$+$	&$+$	&$+$	&$+$	\\
$C_4$ 		&$\pm$	&$+$	&$0$	&$\pm$	&$\pm$	&$\pm$	&$-$	&$\pm$	&$\pm$	 \\
$D_4$ 		&$+$	&$0$	&$0$	&$\pm$	&$-$	&$\pm$	&$\pm$	&$-$	&$-$	 \\
$E_4$ 		&$0$	&$0$	&$0$	&$+$	&$+$	&$+$	&$+$	&$+$	&$+$	 \\
$F_4$	 	&$0$	&$0$	&$0$	&$0$	&$0$	&$0$	&$0$	&$+$	&$+$	 \\ \hline
\end{tabular}
\caption{$\epsilon^4$ coefficients for the case that $\epsilon^2$, and thus also $\epsilon^3$,
 are fine-tuned to vanish; $m_f=0$. }
\label{tab:eps4}
\end{table}

We start with the case where $X$ is charged under $SU(3)$. One immediately
sees that the cases $(S,S)$ and $(V,V)$ are not affected. The problematic ones
are $(S,V)$ and $(V,S)$ which now get a positive $E$ as is the case of $(V,V)$ in
the leading order. Unfortunately, ratios of the other coefficients do not help
if one has no further information on couplings and/or masses of the intermediate
particles. We have marked this possibility in Fig.~\ref{fig:strategy} 
with the yellow boxes surrounded by  red dashed lines. However, we want
to stress that this requires fine-tuning between different couplings which
although being unlikely, cannot be excluded on logical grounds.
 
In the case that $X$ is a $SU(3)$ singlet the situation gets even
a little bit more complicated, because now also in case of $(S,S)$ the
$D$ coefficient is non-zero. However, it is still positive and, thus,
it can for sure not be confused with the case of a new fermion. In the
$(V,V)$ case on the other hand we get in principle even more information 
as now also the $F$ is non-zero which immediately tells us that there is
a special combination of couplings.

\subsection{Final states containing massive  SM-fermions}
\label{subsec:massivestrategy}

Here we summarise the changes for massive SM-fermions, which in practice only
is important for top-quarks.  The signs of the coefficients are
given in Table~\ref{tab:massivedecaycoeff}. It turns out, that  things hardly change 
but for the fact that one has to fit 
more coefficients.

\begin{table}[t]
\centering
\begin{tabular}{|c| c c | cc  | cc | cc | }
\hline
 & \multicolumn{2}{c|}{$(S,S)$} & \multicolumn{2}{c|}{$(S,V)$} & \multicolumn{2}{c|}{$(V,S)$} & \multicolumn{2}{c|}{$(V,V)$} \\ \hline
$\epsilon^2$ & $s$ 	& $c$/1+2	& $s$ 	& $c$/ 1+2 & $s$ & $c$/1+2 & $s$ & $c$/1+2 \\ \hline 
$A_2$ 	     & $0$ 	& $0$      	&$+$	& $+$      &$+$	&$+$               &$+$&$+$        \\
$B_2$ 	     & $\pm$	&$\pm$	   	&$\pm$	& $\pm$    &$\pm$	&$\pm$     &$\pm$&$\pm$       \\
$C_2$ 	     &  $\pm$	&$\pm$	   	&$\pm$	& $\pm$    &$+$	&$+$             	&$\pm$&$\pm$            \\
$D_2$ 	     &    $0$	&$0$	   	&$\pm$  & $\pm$    &$+$	&$+$              &$\pm$&$\pm$             \\
$E_2$ 	     & $0$	&$0$ 	   	&$0$	& $0$      &$0$	&$0$        &     $+$&$+$                 \\
\hline
$\epsilon^3$ &     	 &            	 &          &        &       	     &              &  	          &          \\ \hline 
$A_3$	     & 0   	 &0	  	 &$\pm$     &$\pm$   &$\pm$        &$\pm$          &$\pm$        &$\pm$          \\
$B_3$ 	     & $\pm$ 	 &$\pm$ 	 &$\pm$     &$\pm$   &$\pm$        &$\pm$          &$\pm$        &$\pm$          \\
$C_3$ 	     & $\pm$	 &$\pm$  	 &$\pm$     &$\pm$   &$\pm$        &$\pm$          &$\pm$        &$\pm$          \\
$D_3$ 	     & $ 0$ 	 &0 		 &$\pm$     &$\pm$   &$\pm$        &$\pm$          &$\pm$        &$\pm$          \\
$E_3$ 	     &   0 	 &0   		 &$0$        &0      &0              &0             &$\pm$        &$\pm$          \\
\hline
$\epsilon^4$ & 	        &       	&        &   	   &           	   &         &      &    \\ \hline 
$Z_4$ 	     & 0  	&$0 $ 	   	&$ 0$    &$ 0$          &0           &0             &$-$     &   $-$          \\
$A_4$ 	     & $+$ 	&$0$  		&$+$  	 &$+$          &$+$        &$+$          &$+$     &   $+$          \\
$B_4$ 	     & $\pm$ 	&$+$ 		&$\pm$   &$\pm$          &$\pm$        &$\pm$          &$\pm$     &   $\pm$          \\
$C_4$ 	     & $\pm$ 	&$+$  		&$\pm$   &$\pm$          &$\pm$        &$\pm$          &$\pm$     &   $\pm$          \\
$D_4$ 	     & $\pm$ 	&$-$	   	&$\pm$   &$\pm$          &$\pm$        &$\pm$          &$\pm$     &   $\pm$          \\
$E_4$ 	     &   0  	&0	 	&$\pm$   &$\pm$          &$\pm$        &$\pm$          &$\pm$     &$\pm$             \\
$F_4$ 	     &   0  	&0	 	&0       &0              &$0$        &0              &$\pm$     &$\pm$             \\
\hline
\end{tabular}
\caption{Same as Table~\ref{tab:masslessdecaycoeff} but for massive SM-fermions.}
\label{tab:massivedecaycoeff}
\end{table}

\begin{table}[t]
\centering
\begin{tabular}{|c| c cc | cc  | cc | cc | }
\hline
 & \multicolumn{3}{c|}{$(S,S)$} & \multicolumn{2}{c|}{$(S,V)$} & \multicolumn{2}{c|}{$(V,S)$} & \multicolumn{2}{c|}{$(V,V)$} \\ \hline
$\epsilon^4$ 	& $s$ 	& $c$ & 1+2 & $s$ 	& $c$/1+2 	& $s$ 	& $c$/1+2 & $s$ & $c$/1+2 \\ \hline 
$Z_4$		&$0$	&$0$	&$0$	&$0$	&$0$	&$0$	&$0$	&$-$	&$-$	\\
$A_4$		&$+$	&$0$	&$0$	&$+$	&$+$	&$+$	&$+$	&$+$	&$+$	\\
$B_4$ 		&$\pm$	&$+$	&$+$	&$\pm$	&$\pm$	&$\pm$	&$+$	&$\pm$	&$\pm$	\\
$C_4$ 		&$\pm$	&$+$	&$+$	&$\pm$	&$\pm$	&$\pm$	&$-$	&$\pm$	&$\pm$	 \\
$D_4$ 		&$+$	&$0$	&$0$	&$\pm$	&$\pm$	&$\pm$	&$\pm$	&$\pm$	&$\pm$	 \\
$E_4$ 		&$0$	&$0$	&$0$	&$\pm$	&$+$	&$+$	&$+$	&$\pm$	&$+$	 \\
$F_4$	 	&$0$	&$0$	&$0$	&$0$	&$0$	&$0$	&$0$	&$+$	&$+$	 \\ \hline
\end{tabular}
\caption{Same as Table \ref{tab:eps4} but $m_f\ne 0$.}
\label{tab:massiveeps4}
\end{table}

Comparing tables~\ref{tab:masslessdecaycoeff} and \ref{tab:massivedecaycoeff}
one sees that the same strategy can be used in principal. However, for distinguishing
between the $(S,V)$ and $(V,S)$ cases the ranges for the ratios of the coefficients
change. Moreover, only in the ratio $D/C$ the unknown couplings cancel
and we find
\begin{equation}
\begin{array}{rcl}
(S,V): & D/C =&\frac{(\tau_Y-1)^2-4 \tau_f^2}{12 \tau_f^2+22
   \tau_Y^2-4 \tau_Y-2} \,\,\,[-\infty,-\frac{1}{3}] \cup [0,\infty] \\
(V,S): & D/C =& \frac{(\tau_Y-1)^2-4 \tau_f^2}{2 \left(6 \text{$\tau
   $f}^2-\tau_Y (\tau_Y+2)+11\right)} \,\,\,[0,\frac{1}{22}] 
\end{array}
\label{eq:tau}
\end{equation}
where one has to use 
\begin{equation}
1\geq \tau_Y+2 \tau_f
\label{eq:energycons}
 \end{equation}
due to total energy/momentum conservation. As in the case of massless SM-fermions,
the overlap region of the two intervals is for the case $\tau_Y \to 1 - 2 \tau_f$,
e.g.~the kinematical limit, where all particles are practically at rest in the centre of
mass system of $X$. In general
this ratio will be either negative or much larger than 1/2 in the $(S,V)$ case. 
Note that we have $A=0$ for the $(S,S)$ case which, thus, serves as a 
confirmation of this case. Also in case of a new fermion we arrive at the same
conclusions because
\begin{eqnarray}
\mathrm{sign}(A_4)=+ \hspace{3mm},\hspace{3mm}
\mathrm{sign}(B_4)=\pm \hspace{3mm},\hspace{3mm}
\mathrm{sign}(C_4) = \pm \hspace{3mm},\hspace{3mm}
\mathrm{sign}(D_4) = - \hspace{3mm},\hspace{3mm}
\mathrm{sign}(E_4)=0 \nonumber \\
\label{eq:signMFFM2}
\end{eqnarray}
The only exception is where this decay is mediated solely by scalars in the
third topology as in this case $A_4=0$.

For completeness, we also give the results in the case that the leading
orders vanish in the case of decays of bosons in Table~\ref{tab:massiveeps4}.
It turns out that this case is the same as for the case of massless SM-fermions
discussed above except that now in general also $A$ will be non-zero.

\subsection{Dependence on the mass of the intermediate particle(s)}

\begin{figure}[t]
\centering
\begin{picture}(200,470)
\put(-120,320){\includegraphics[width=7cm]{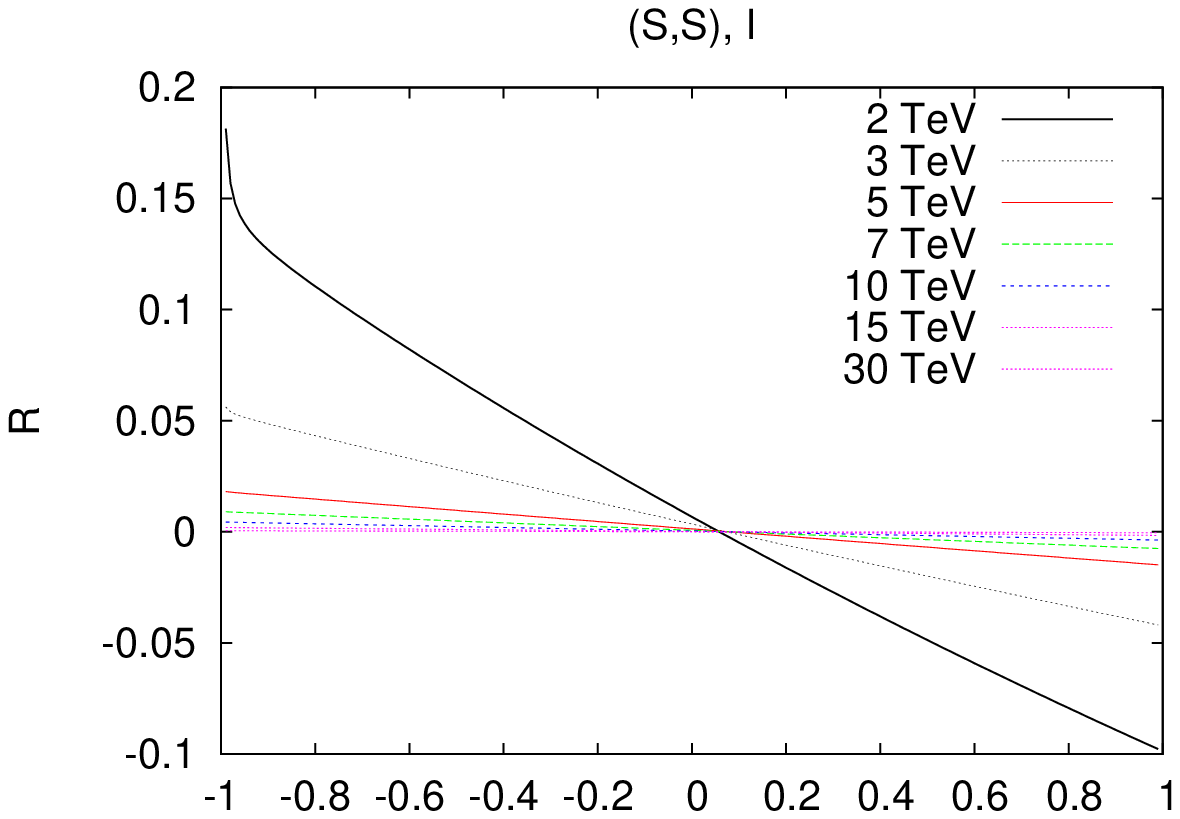}}
\put(-20,310){$\hat{s}$}
\put(100,320){\includegraphics[width=7cm]{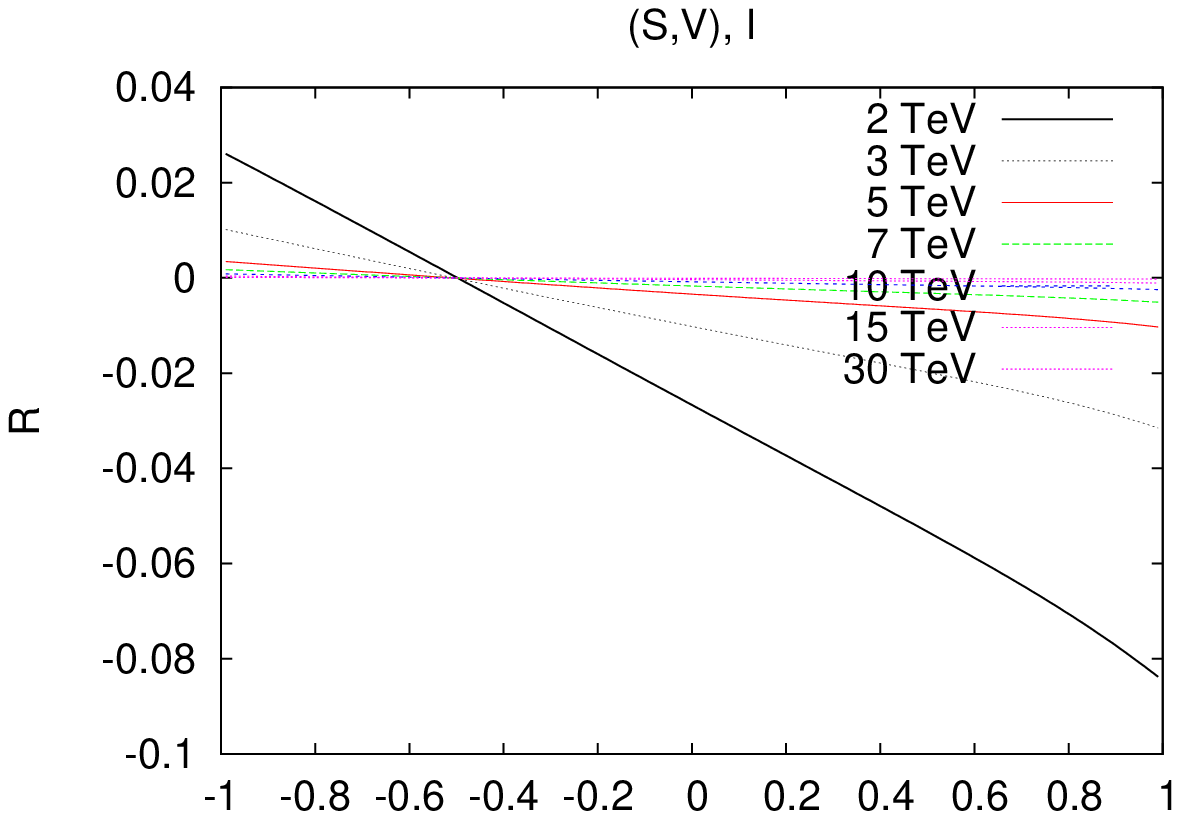}}
\put(220,310){$\hat{s}$}
\put(-120,160){\includegraphics[width=7cm]{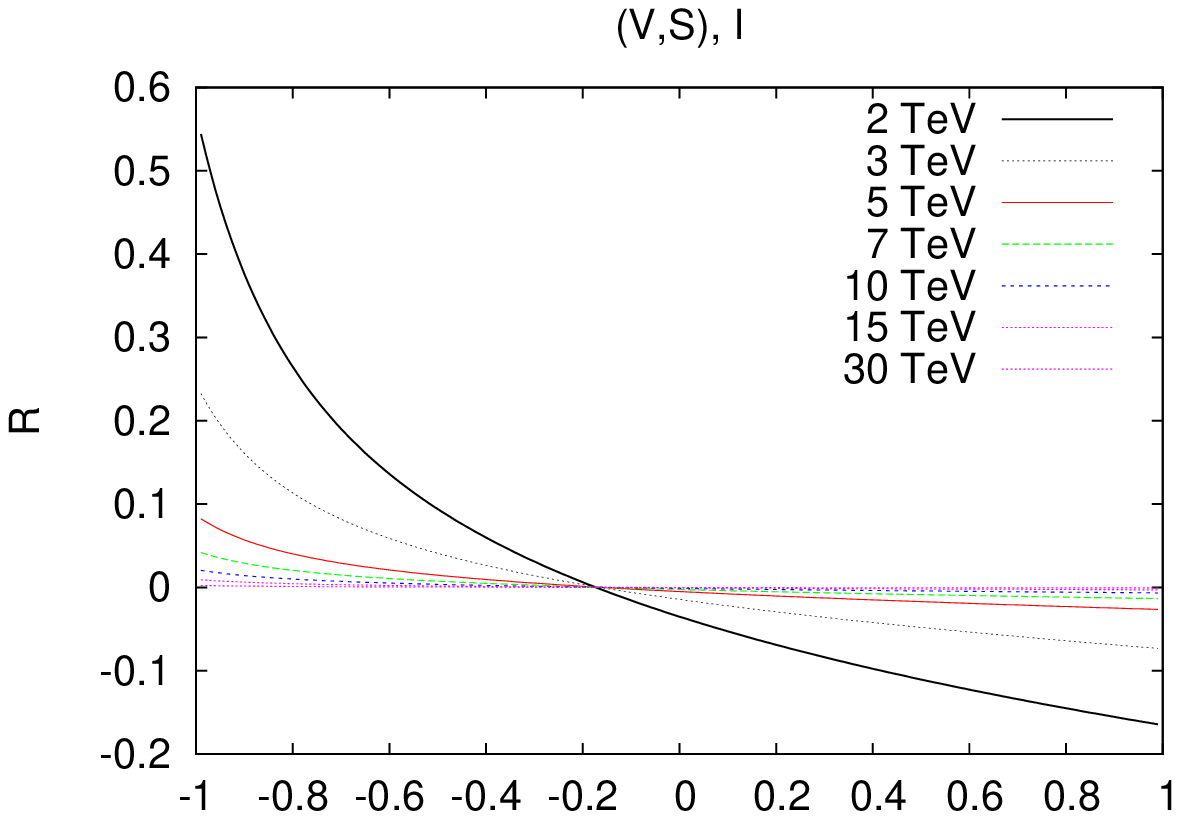}}
\put(-20,150){$\hat{s}$}
\put(110,160){\includegraphics[width=7cm]{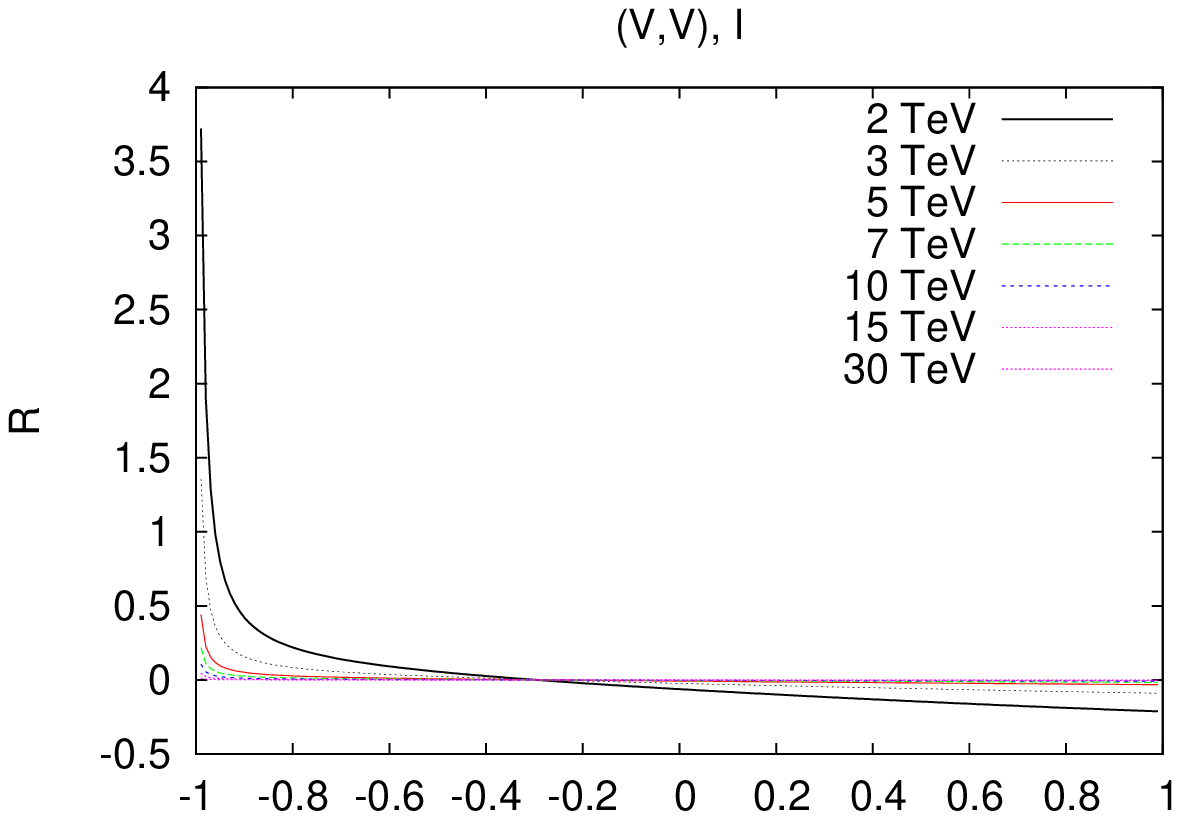}}
\put(220,150){$\hat{s}$}
\put(-120,5){\includegraphics[width=7cm]{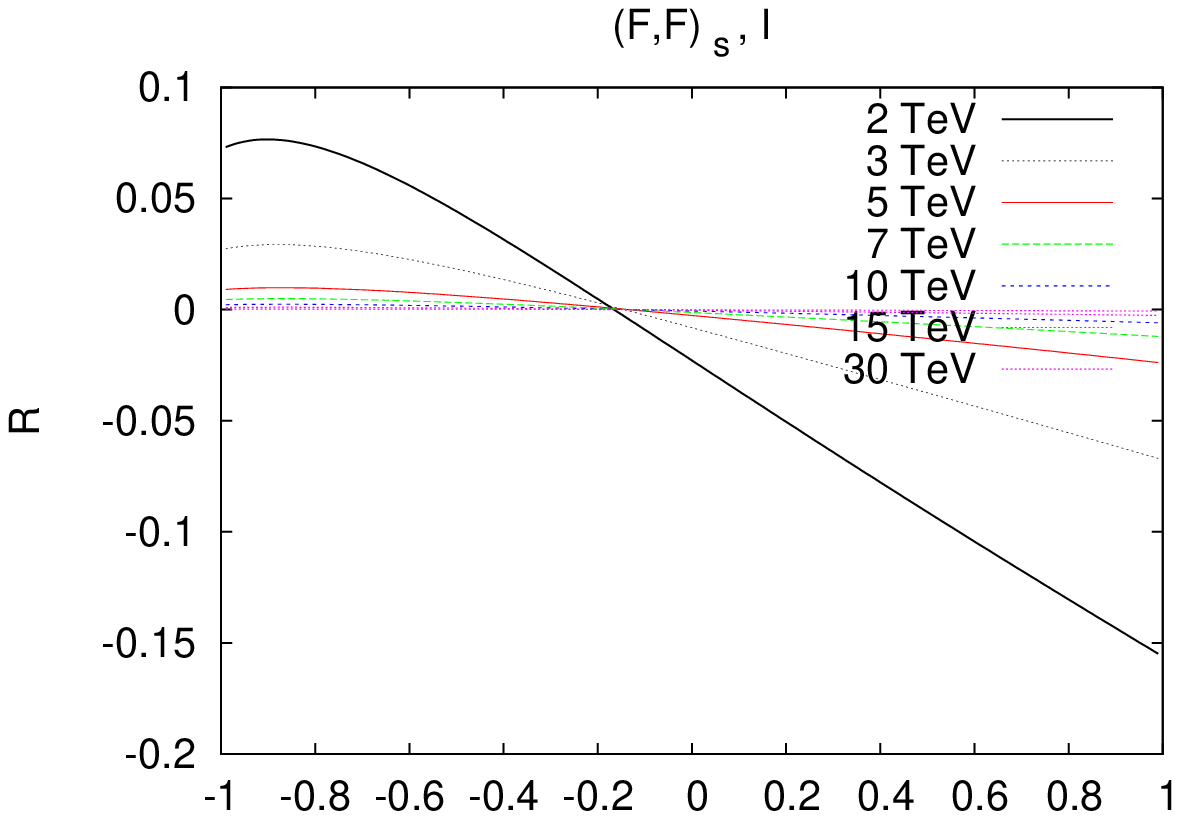}}
\put(-20,-5){$\hat{s}$}
\put(110,5){\includegraphics[width=7cm]{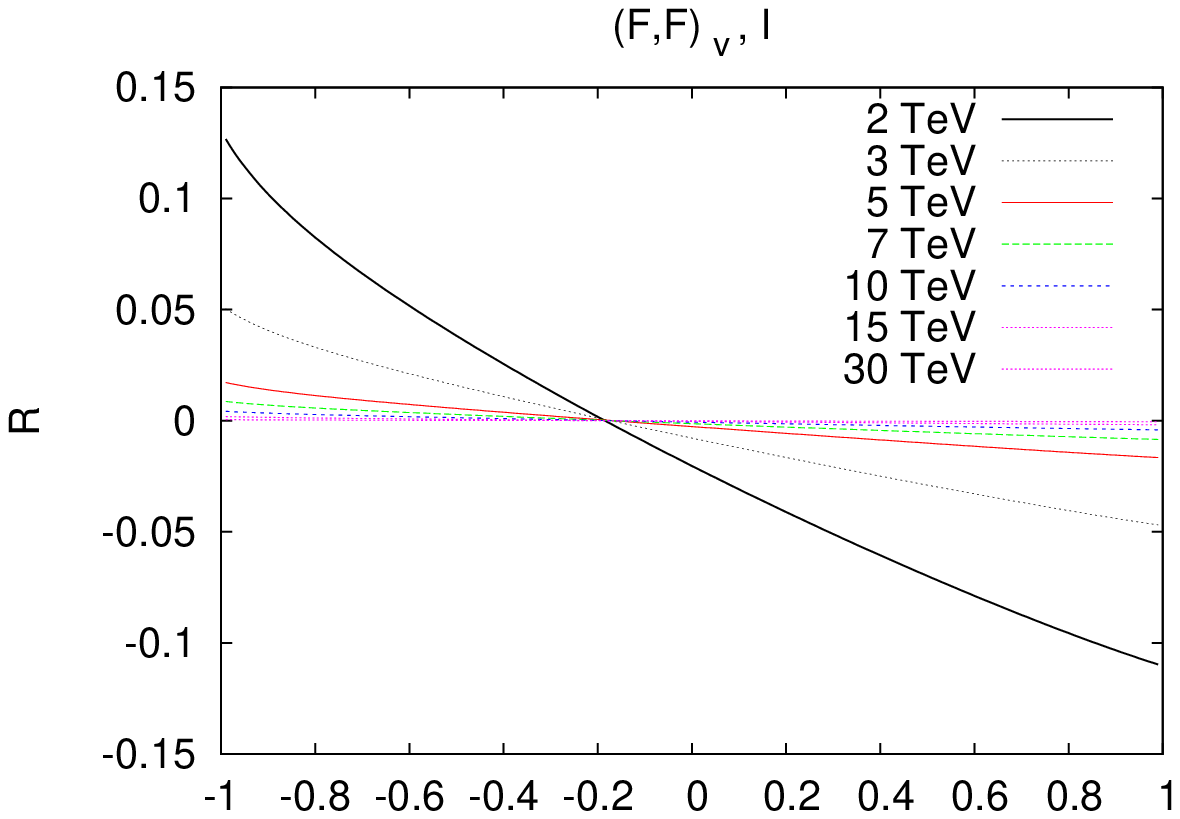}}
\put(220,-5){$\hat{s}$}
\end{picture}
\caption{Ratio $R$ for  the processes  
$X\rightarrow f \bar{f} Y$ for scenario (i) taking $m_X=1 \,\,TeV$, $m_f=0$, $m_Y=0.1\,\, TeV$.
We have calculated the decay width for the following masses of the intermediate particle:
 $m_I=2,3,5,7,10,15,30 \,\,TeV$.}
\label{fig:rhoI}
\end{figure} 

We now address the question how small $\epsilon$ has to be so that our
strategy works. For this we consider two examples: (I) $g(r)=g(l)=n(r)=n(l)=1$
where the leading order dominates and (II) $g(r)=n(r)=1$ but
$g(l)=n(l)=0$ so that the leading order vanishes in  case of the bosonic decays
 and the subleading
orders become dominant. Note, that in case of new fermions we did not manage
to find a combination where the leading order vanishes. In all cases we have 
taken $m_X=1~TeV$, $m_Y=100~GeV$ and
$m_f=0$. We have checked that our results do not depend crucially on these
values except for the cases where $m_Y$ gets close to $m_X$ which would imply soft SM
fermions and experimental difficulties to observe the decay.

\begin{figure}[!t]
\centering
\begin{picture}(200,470)
\put(-120,320){\includegraphics[width=7cm]{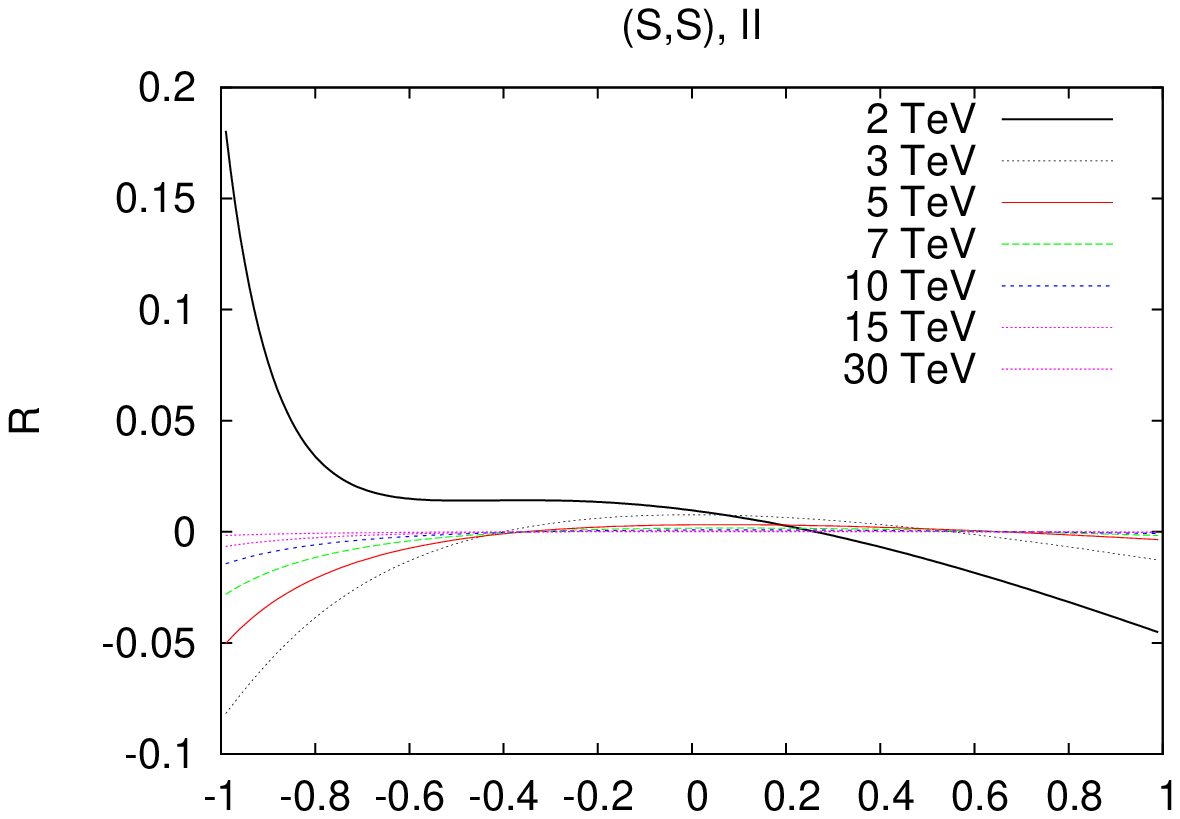}}
\put(-20,310){$\hat{s}$}
\put(110,320){\includegraphics[width=7cm]{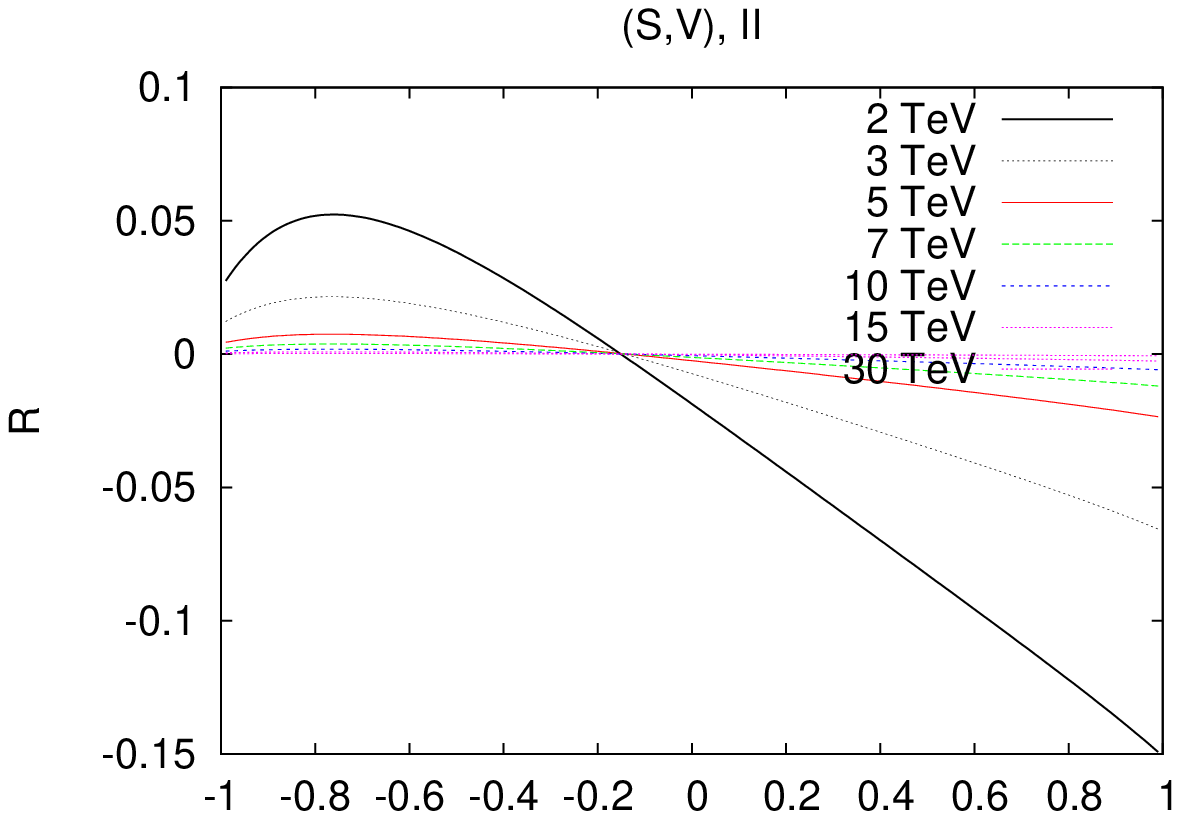}}
\put(220,310){$\hat{s}$}
\put(-120,160){\includegraphics[width=7cm]{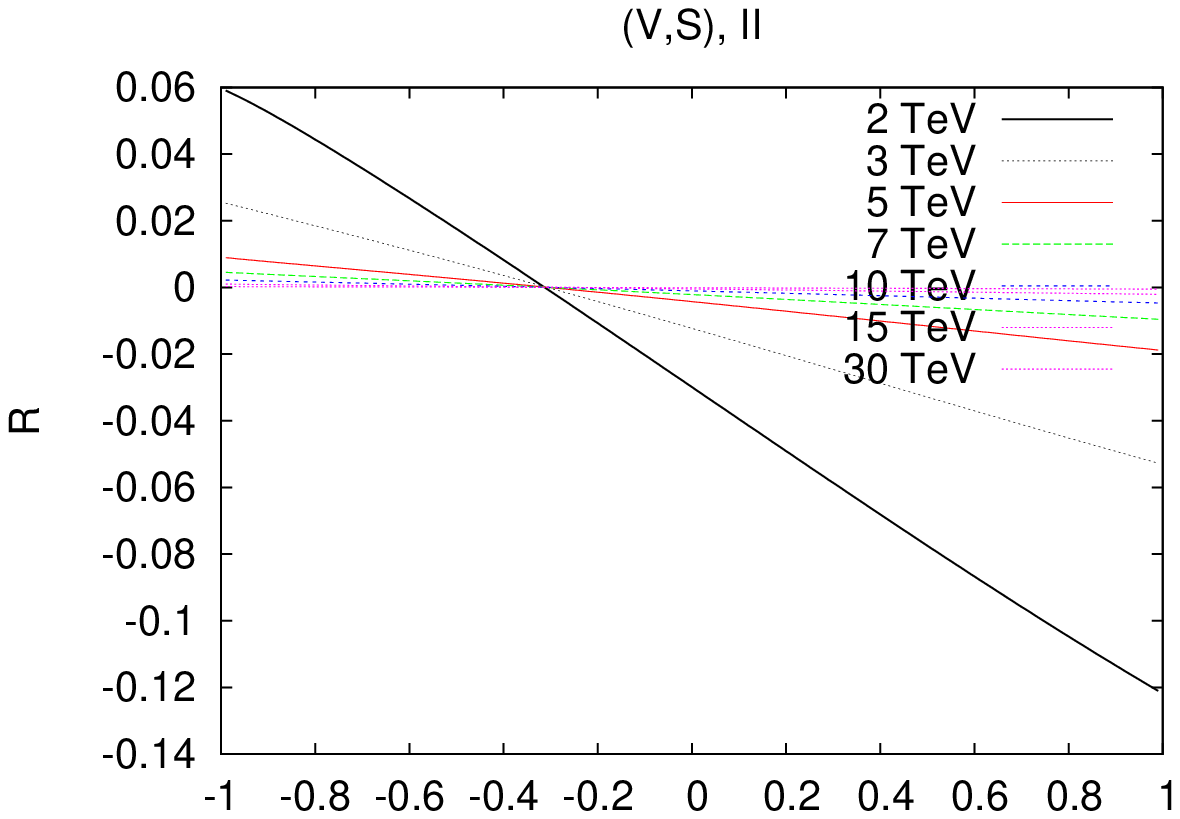}}
\put(-20,150){$\hat{s}$}
\put(110,160){\includegraphics[width=7cm]{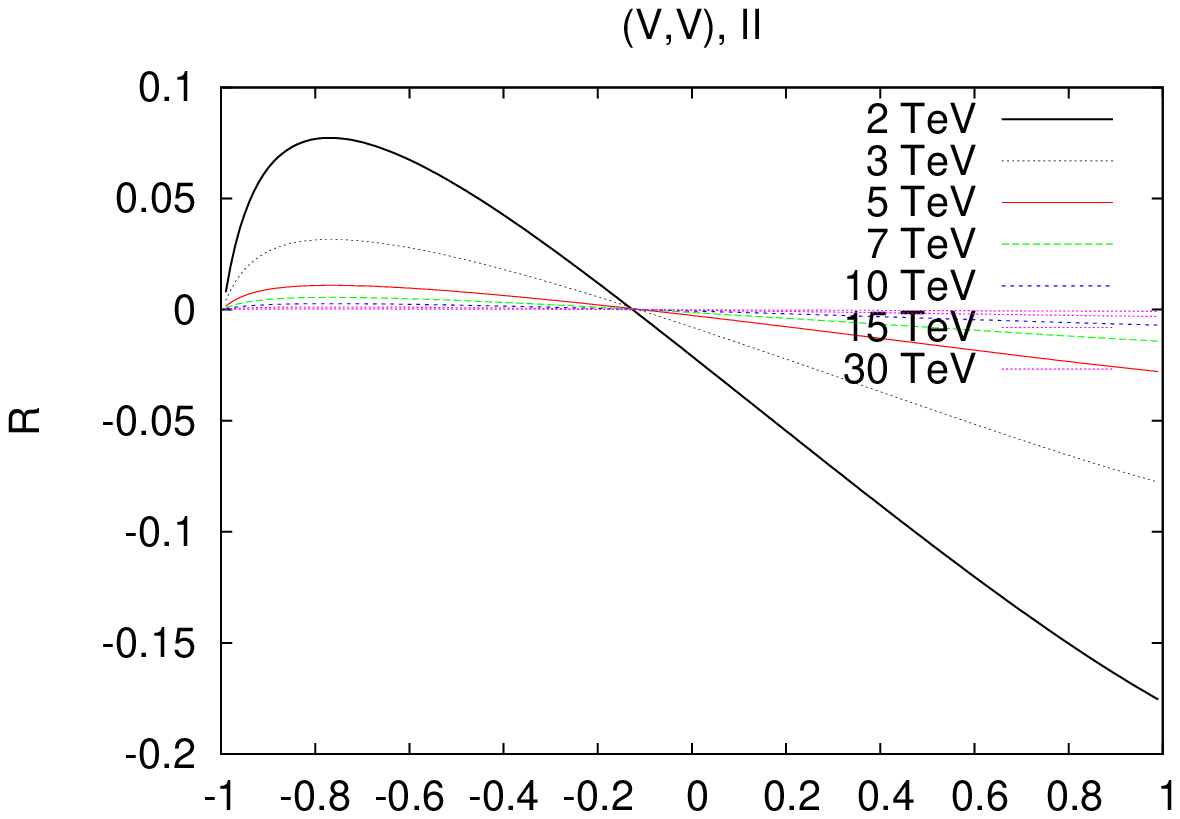}}
\put(220,150){$\hat{s}$}
\put(-120,5){\includegraphics[width=7cm]{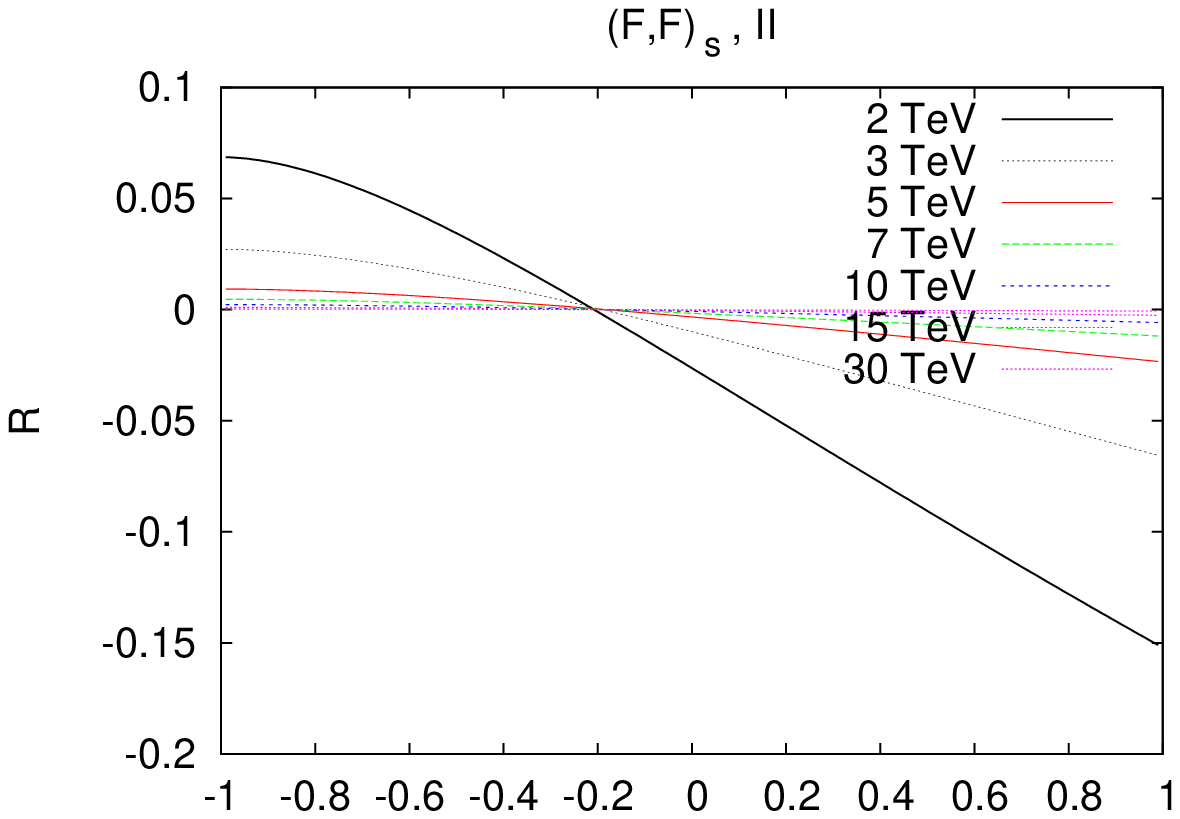}}
\put(-20,-5){$\hat{s}$}
\put(110,5){\includegraphics[width=7cm]{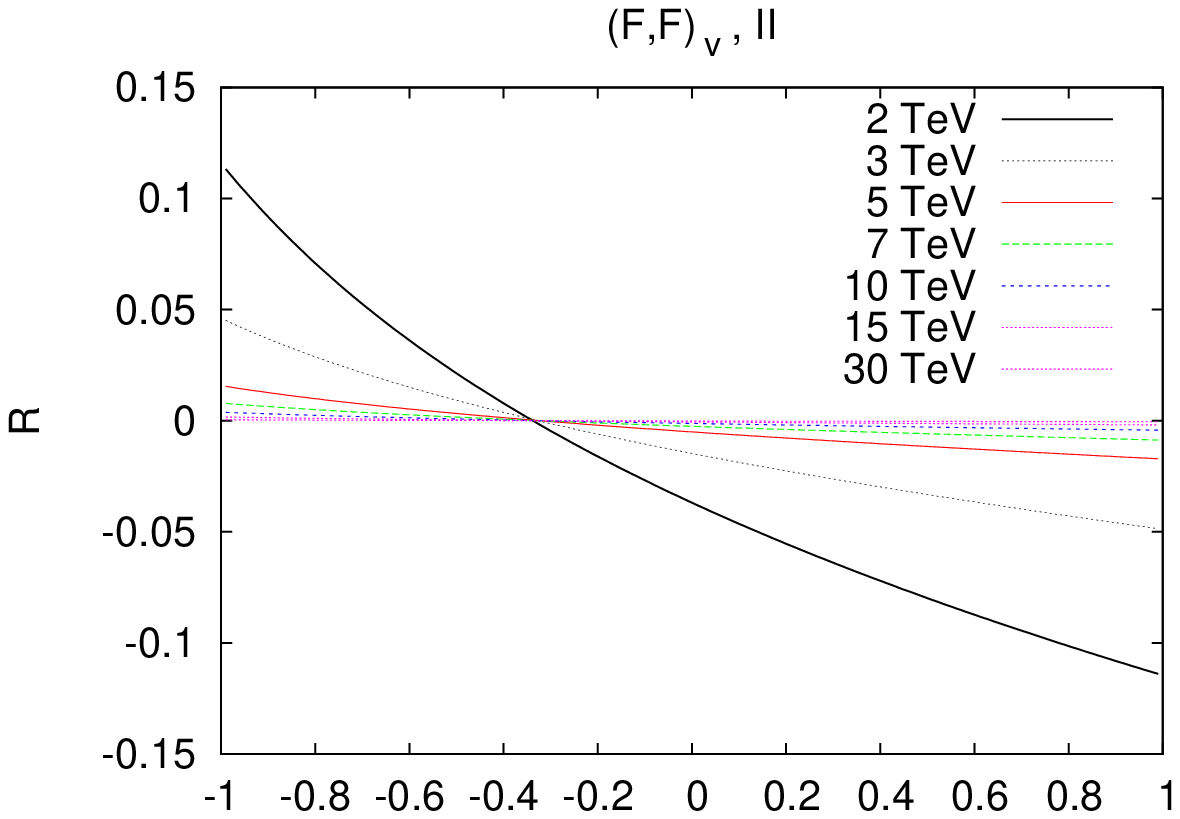}}
\put(220,-5){$\hat{s}$}
\end{picture}
\caption{
Relative deviation for differential decay width the processes  $X\rightarrow f \bar{f} Y$ with  couplings (II).
}
\label{fig:rhoII}
\end{figure}

 In figures \ref{fig:rhoI} and \ref{fig:rhoII} we show the relative deviation
\begin{eqnarray}
R & = & \frac{d\, \Gamma_\epsilon -d\, \Gamma_H}{d\, \Gamma_H} 
\hspace{5mm} \mathrm{with} \hspace{5mm} 
d \Gamma_i = \frac{1}{\Gamma_i} \frac{d\, \Gamma_i}{d\, \hat s} \; ,
\end{eqnarray}
and $H$ denotes the limit $m_I \to \infty$ and $\Gamma_\epsilon$ the differential width for
a given $\epsilon$. We find that for a decaying scalar and for a decaying
fermion the deviation
is always below 20\%. In case of a decaying vector particle the situation is more difficult
and only for $\epsilon \le 1/5$ we get $R \lsim 0.2$ for all values of $\hat s$. The reason
for these large deviations for $|\hat s|$ close to one is, that here the differential widths
becomes zero and the rise/fall at the ends of the interval gets steeper the smaller 
$\epsilon$ is. This also implies, that in the corresponding intervals for $\hat s$ one
will observe only a few events. The situation improves for a decay of a vector boson
if the subleading
terms become dominant as in Fig.~\ref{fig:rhoII}.

In summary: our strategy should work well in all cases if $\epsilon$ is below $1/5$,
in cases of scalar particles or spin 1/2 fermion  $\epsilon =1/2$
 is already a reasonable value. This is for example
a natural value for gluino decays in supersymmetric models.


\section{Testing our Strategy with Monte Carlo Simulations}
\label{sec:montecarlo}
We now test our strategy in Monte Carlo simulations at the parton
level.  For this one of us created data sets and the second one tried
to find out the spin-assignments without any prior knowledge but
 $m_Y$ and $(m_X-m_Y)$ and their uncertainties. We presume that the first one
is given by an independent source with a precision of $10\%$ and the
second one with a precision of $3\%$. For definiteness we have taken
$m_X=1 ~TeV$, $m_Y=100 ~GeV$ and $m_f=0$.  In the following we will
denote by $(B,C)$, $(B,C,D)$, $(B,C,D,E)$ and $(B,C,D,E,F)$ the
differential width in Eq.~(\ref{eq:decayrate}) where all but the given
coefficients are zero.

\subsection{Fitting procedure}

In practice one will not have $\frac{d\Gamma}{d\hat s}(\hat s)$ but one will have
the number of events for a given interval $[\hat s_i,\hat s_i + \Delta \hat{s}]$.
For this reason we actually fit 'distributions' of the form
\begin{equation}
\sum_{i=1}^n \int_{-1+(i-1) \Delta \hat s}^{-1+i \Delta \hat{s}} 
\frac{d \Gamma}{d \hat s}d \hat s
\end{equation}
where $\Delta \hat{s} = 2/n$, $n$ is the number of bins considered and 
$\frac{d \Gamma}{d \hat s}$ is given by Eq.~(\ref{eq:decayrate}). 
For the creation of the 'data' we have used our model file 
 for generic particles and couplings \cite{LAMAmodel}  for the
O'Mega/WHIZARD Monte Carlo generator \cite{Kilian:2007gr,Moretti:2001zz}
which contains generic particles and couplings.

For fitting we use a linear least squares approach as described e.g. in \cite{Deuflhard}
 and the references given therein.
We  will exemplify this  for the case of $(B,C,D)$:
\begin{equation}
\left(\begin{array}{ccc}
x_1 & x_1^2 & x_1^3 \\
x_2 & x_2^2 & x_2^3 \\
 &\vdots & \\
x_n & x_n^2 & x_n^3
\end{array}\right)
\cdot 
\left(
\begin{array}{c}
B\\
C\\
D
\end{array}
\right)=
\hat X \cdot \left(
\begin{array}{c}
B\\
C\\
D
\end{array}
\right)=
\left(
\begin{array}{c}
\mbox{data}_1\\
\mbox{data}_2\\
\vdots\\
\mbox{data}_n
\end{array}
\right)
\label{eq:fit1}
\end{equation}
where $n$ is the number of bins. This equation can be solved by rewriting it as:
\begin{equation}
(\hat{X})^T\hat{X} \cdot \vec{c}=(\hat{X})^T\vec{d}
\label{eq:fit2}
\end{equation}
where $\vec{c} = (B,C,D)^T$  and $\vec{d}$ contains the 'measured' data 
of the differential width
integrated over intervals of length $\Delta \hat s$.
There are various methods to solve this equation, e.g.\ QR-decomposition. The fit is reliable
if the matrix  $(\hat{X})^T\hat{X}$ is well-conditioned, e.g.~if its eigenvalues are
of similar order of magnitude.  
After solving Eq.~(\ref{eq:fit2}) we
calculate the corresponding $\chi^2$
\begin{eqnarray}
\chi^2&=& \frac{1}{n-j}\sum_i \frac{(\mbox{Expected}_i-\mbox{Observed}_i)^2}{\mbox{Expected}_i}
\label{eq:chisquared}
\end{eqnarray}
where $n$ is the number of bins (= number of data points) and $j$ the degrees of
freedom (in this case the number of coefficients) of the fit function.
We estimate the error for the coefficients by adding a Poisson noise 
to the data and determine the corresponding confidence interval after fitting.

\subsection{A supersymmetric example}

As a first test we study a focus point scenario which is inspired by
SPS2 \cite{Allanach:2002nj}: $m_0= 3\cdot 10^3$ GeV, $m_{1/2} = 3\cdot
10^2$, $A_0 =0$, $\tan\beta=10$ and sign$(\mu)>0$. We simulate for the
gluino decay 2k and 10k events and study also the effect of different
binning sizes, namely 10 and 50 bins.  Afterwards we fit the resulting
distribution to all possible cases. Here we have assumed that the
following information on the masses is given: $(m_X-m_Y)=688\pm 23~GeV$,
 $m_Y=121\pm 12~GeV$ with a Gaussian distribution.  The results
are summarised in Table~\ref{tab:fitmffm}. As one does not know neither
the absolute values of the couplings nor the intermediate masses, one
has the freedom to normalise $B$ to 1 and, thus, only the other
coefficients and their uncertainties are given.  The $\chi^2$ favours
for both 10 and 50 bins slightly the $(B,C,D)$ polynomial.  However,
if higher powers in $\hat s$ are included, one still obtains a good
fit and the $\chi^2$ will not be sufficient to discriminate between
the different possibilities. This is a quite generic feature because
usually there is a hierarchy between the non-zero coefficients:
$|B|,|C| \gg |E|,|F|$.  The $(B,C)$ case can be ruled out since $B\neq
C$ within the range of the error bars. The smallness of the parameters
and the large errors on the $E$ and $F$ coefficients of the
$(B,C,D,E)$ and $(B,C,D,E,F)$ models suggest that those values are
zero. The remaining model is $(B,C,D)$ with negative $D$ which suggests
$(F,F)$. We find it encouraging that one gets already for 2000 events
first information including that the $(S,S)$ case can be excluded.

\begin{table}[t]
 \centering
\begin{tabular}{|c |cr|cr|cr|}
\hline
 Coef.	& $\chi^2$ 	& (F,F)$_{10}$  & $\chi^2$& (F,F)$_{50}$ 	& $\chi^2$& (F,F)$_{10,2k}$	\\ \hline
	 C	& 5.9	&$ 0.210 \pm  0.061   $	&1.50	&$0.208\pm0.071  $   	&1.58	&$0.208\pm 0.126  $			\\ \hline
 C	& 0.55	&$0.089 \pm  0.077  $	&0.62	&$0.089\pm  0.076  $		&1.55	&$0.089\pm  0.155  $		\\ 
	 D	& 	&$-0.232 \pm   0.106 $	&	&$-0.227 \pm  0.110 $ 	&	&$-0.227\pm   0.252  $ \\ \hline
 C 	& 0.63	&$ 0.079 \pm  0.141  $	&0.63	&$0.067  \pm 0.128   $	&1.75	&$ 0.067\pm  0.300   $  		\\ 
	D	& 	&$ -0.218\pm  0.150   $	&	&$-0.197  \pm   0.139 $	&	&$-0.197 \pm   0.322  $		\\ 
	E	& 	&$0.028 \pm   0.227   $ 	&	&$0.060 \pm  0.228 $ 		&	&$0.060 \pm   0.536   $	\\ \hline
 C	& 0.75	&$0.080 \pm   0.149 $  	&0.64	&$0.077 \pm  0.160  $  	&1.14	&$ 0.077\pm  0.401  $		\\ 
	D	& 	&$-0.215 \pm  0.456   $	&	&$-0.169\pm0.340   $ 	&	&$ -0.169\pm  1.108   $		\\ 
	 E	& 	&$0.025 \pm   0.260 $  	&	&$0.035\pm 0.311  $   	&	&$ 0.035\pm  0.738    $			\\ 
	 F	& 	&$ -0.006\pm  0.516  $ 	&	&$-0.046\pm 0.443    $		&	&$-0.046\pm 1.315  $ 		\\ \hline
\end{tabular}
\caption{Testing of the SUSY focus point taking 10 bins in case of 2000 events
and 10 and 50 bins for 10000 events. Input for the fit is: 
$(m_X-m_Y)=688\pm 23~GeV$, $m_Y=121\pm 12~GeV$  with gaussian 
distribution, $m_I=3026~GeV$. 
The coefficients are normalised such that $B=1$ and the uncertainties are at $3 \sigma$. 
The analytic values for 
the coefficients are $B=1$, $C=0.123$, $D=-0.188$. }
\label{tab:fitmffm}
\end{table}

\subsection{Large sample tests}

In the second step we have tested our strategy for a large
set of random couplings fixing however the kinematics to
$m_X = 1~TeV$, $m_Y=100~GeV$ and all $m_I= 15~TeV$. The latter 
number is not crucial as long as it is above $5~TeV$ ($2~ TeV$) in case
of decaying vector bosons (decaying scalars and fermions). 
 We have generated
100 different sets with random couplings for each of 
the decays $(S,S)$, $(S,V)$, $(V,S)$, $(V,V)$ and $(F,F)$ 
and we have generated for each set of couplings $10^4$, $10^5$ and
$10^6$ events. In an ideal world one could use the strategy depicted
in Fig.~\ref{fig:strategy} without any problems. In reality there will
be some smearing of the data from the measurement itself as well as from
the background subtraction.
To be sure that we do not miss anything we have slightly advanced
our strategy and apply it to the following model
list
\begin{eqnarray}
\{(S,S),(S,V),(V,S),(F,F),(V,V),(V,V)_4\}
\label{eq:list}
\end{eqnarray}
and the  corresponding differential widths. $(V,V)_4$ denotes the case, where 
only the 4th order in $\epsilon$ of (V,V) remains.  We 
have tested for the following
criterions:
\begin{enumerate}
 \item\label{item:bneqc}  $B\neq C$ $\rightarrow$ remove $(S,S)$
\item\label{item:dneg} $D >0$ $\rightarrow $ remove $(F,F)$; $D <0$ $\rightarrow $ remove $(S,V),(V,S)$
\item\label{item:cbsv} C/B in (S,V) interval $\rightarrow $ if not remove $(S,V)$
\item\label{item:cbvs} C/B in (V,S) interval $\rightarrow $ if not remove $(V,S)$
\item\label{item:dbsv} D/B in (S,V) interval $\rightarrow $ if not remove $(S,V)$
\item\label{item:dbvs} D/B in (V,S) interval $\rightarrow $ if not remove $(V,S)$
\item\label{item:dcsv} D/C in (S,V) interval $\rightarrow $ if not remove $(S,V)$
\item\label{item:dcvs} D/C in (V,S) interval $\rightarrow $ if not remove $(V,S)$
\item\label{item:epos} if $E<0$ in $(V,V)/(V,V)_4$, $\rightarrow $ remove $(V,V)/(V,V)_4$
\item\label{item:fpos} if $F<0$ in $(V,V)_4$, $\rightarrow$ remove $(V,V)_4$
\item\label{item:chi} Optional: Remove all models with $\chi^2>3$
\item\label{item:efsmall} Optional: Remove $(V,V)/(V,V)_4$ if $E,F <0.001$ respectively 
\end{enumerate}
Every time one criterion could be applied or not fulfilled within the
range of the error bars, the corresponding model was cancelled from the
list in Eq.~(\ref{eq:list}).

In table \ref{tab:fits1} we have summarised our results for these
different Monte Carlo data sets where we give the number of 
the  remaining models after going through the different
criteria. The first row e.g. means, that we started with 100 different
data sets for the $(S,S)$ decay, applied our tests and after that, 100 $(S,S)$
models remained, 0 of $(S,V)$, 20 of $(V,S)$, 100 of $(F,F)$ and so on.  There were
no data sets where only a wrong model remained.

\begin{table}[ht]
\centering
\begin{tabular}
{|c|c ccccc|}
\hline
	 &	(S,S)	&	(S,V)	&	(V,S)	&	(F,F)	&	(V,V)	&	(V,V$_4$)	\\ \hline
\multicolumn{7}{|l|}{ $10^4$ events:}							\\ \hline 
(S,S): 	&	 \bf 100 &	0	&	 20	&	 100	&	 100	&	 100		  \\ 
(S,V): 	&	 0	&	\bf99	&	 0	&	 1	&	 100	&	 99	 	  \\ 
(V,S): 	&	 0	&	0	&	 \bf99	&	 99	&	 100	&	 98	 	  \\ 
(F,F): 	&	 0	&	5	&	 0	&	 \bf100	&	 99	&	 99	 	  \\ 
(V,V): 	&	 0	&	66	&	 0	&	 78	&	 \bf100	&	 100	 	  \\ \hline 
\multicolumn{7}{|l|}{$10^5$ events:}							\\ \hline 
(S,S): 	&	 \bf 100&	0	&	 0	&	 100	&	 100	&	 100		  \\ 
(S,V): 	&	 0	& 	\bf99	&	 0	&	 0	&	 100	&	 100	 	  \\ 
(V,S): 	&	 0	&	0	&	 \bf100	&	 100	&	 100	&	 100	 	  \\ 
(F,F): 	&	 0	&	0	&	 0	&	 \bf100	&	 100	&	 100	 	  \\ 
(V,V): 	&	 0	&	66	&	 0	&	 78	&	 \bf100	&	 100	 	  \\ \hline 
\multicolumn{7}{|l|}{$10^6$ events:}							\\ \hline 
(S,S): 	&	 \bf100 &	0	&	 0	&	 100	&	 100	&	 100		  \\ 
(S,V): 	&	 0	& 	\bf98	&	 0	&	 0	&	 99	&	 100	 	  \\ 
(V,S): 	&	 0	&	0	&	 \bf100	&	 100	&	 100	&	 100	 	  \\ 
(F,F): 	&	 0	&	0	&	 0	&	 \bf100	&	 100	&	 100	 	  \\ 
(V,V): 	&	 0	&	10	&	 0	&	 61	&	 \bf100	&	 100	 	  \\ \hline 
\end{tabular}
\caption{
Number for the remaining models for 100 input models each where the various 
criterions are applied
using $3 \sigma$ uncertainties on the coefficients.
The masses are chosen as $m_Y=100\pm 10~GeV$ and the mass difference $(m_X-m_Y)=900\pm 30~GeV$.
The bold numbers are the correct model fits.} 
\label{tab:fits1}
\end{table}

\begin{table}[ht]
\centering
\begin{tabular}
{|c|c ccccc|}
\hline
	 &	(S,S)	&	(S,V)	&	(V,S)	&	(F,F)	&	(V,V)	&	(V,V$_4$)	\\ \hline
\multicolumn{7}{|l|}{$10^4$ events:}							\\ \hline 
(S,S): 	&	\bf 99/100&	0/0	&	 19/20	&	 99/100	&	 99/85	&	 97/90		  \\ 
(S,V): 	&	 0/0	&	\bf96/96	&	 0/0	&	 0/1	&	 93/95	&	 90/89	 	  \\ 
(V,S): 	&	 0/0	&	0/0	&	 \bf97/97	&	 97/99	&	 99/88	&	 96/81	 	  \\ 
(F,F): 	&	 0/0	&	5/5	&	 /00	&	\bf 98/100	&	 98/89	&	 97/87	 	  \\ 
(V,V): 	&	 0/0	&	46/66	&	 0/0	&	 31/78	&	\bf 95/95	&	 92/90	 	  \\ \hline 
 \multicolumn{7}{|l|}{$10^5$ events:}							\\ \hline 
(S,S): 	&	\bf100/99&	0/0	&	 0/0	&	100/100	&	 97/63	&	 98/63		  \\ 
(S,V): 	&	 0/0	& \bf99/95	&	 0/0	&	 0/0	&	 97/81	&	 94/72	 	  \\ 
(V,S): 	&	 0/0	&	0/0	&\bf100/100	&	100/100	&	 100/66	&	 99/64	 	  \\ 
(F,F): 	&	 0/0	&	0/0	&	 0/0	&\bf99/100	&	 98/66	&	 98/61	 	  \\ 
(V,V): 	&	 0/0	&	46/66	&	 0/0	&	 31/78	&\bf95/95	&	 92/90	 	  \\ \hline
 \multicolumn{7}{|l|}{$10^6$ events:}							\\ \hline 
(S,S): 	&	 \bf100/100 &	0/0	&	 0/0	&	 100/99	&	  100/14&	  96/23		  \\ 
(S,V): 	&	 0/0	& 	 95/98	&	 0/0	&	 0/0	&	 93/54	&	 90/33 	 	  \\ 
(V,S): 	&	 0/0	&	0/0	&\bf99/100 	&	 99/100	&	  98/17	&	 98/28 	 	  \\ 
(F,F): 	&	 0/0	&	0/0	&	 0/0	&\bf100/100 &	  98/22	&	  97/25	 	  \\ 
(V,V): 	&	 0/0	&	 0/10	&	 0/0	&	 0/61 	&	 \bf93/100 	&	 97/64 	 	  \\ \hline
\end{tabular}
\caption{Same as Table~\ref{tab:fits1} but taking into account
either the optional criterion 11 ( exclude $\chi^2>3$) or
criterion 12 (small $E$,$F$) corresponding to the first and second number given at
the various entries.} 
\label{tab:fits2}
\end{table}

The obtained results can be understood as follows: (i) It is easier
to fit a polynomial which has low powers of $\hat s$ by a higher
order polynomial if there is smearing than vice versa. 
(ii) The number of criterions depend on the decay mode, e.g.~it
is easier to exclude $(S,V)$ where 3 criterions are at hand than
$(V,V)$ where only one exists.
(iii) The modulus of the coefficients $E$ and $F$ is usually
up to two orders of magnitude smaller than the modulus of the other
coefficients but the absolute uncertainty is roughly the same for all
coefficients. In particular the third item implies that it will
be rather difficult to exclude a positive $E$ and $F$ in practice
if only one decay channel is considered.

Additionally we have also looked at the optional criteria
item~\ref{item:chi} and~\ref{item:efsmall} separately and
give the resulting numbers in Table~\ref{tab:fits2}. 
The first number of 99/100 (as e.g. in the $(S,S)$ case)
denotes the number remaining after applying the $\chi^2$ criterion,
the second number the same with the small $E$,$F$ criterion.  The results
are (i) the $\chi^2>3$ criterion is most useful if the underlying decay is
$(V,V)$/$(V,V)_4$, since the polynomials with a lower order have a large
$\chi^2$. This is e.g. reflected in the $(V,V)$ decay with 10k events,
and the fitted $(S,V)$ and $(F,F)$ polynomials. Here the number of remaining
processes are reduced by applying the $\chi^2$ test from 66 ($(S,V)$, 
Table~\ref{tab:fits1}) to 46 (Table~\ref{tab:fits2}) and for $(F,F)$ from
 78 to 31.  (ii) The criterion for small $E,F$
works very good for high statistics since the fit gives values close to 0.
 This is reflected in the last row of
Table~\ref{tab:fits2}, where the underlying process is $(F,F)$. After
applying of the common criterions, 100 of $(V,V)$ and 100 of $(V,V)_4$
models remain (see Table~\ref{tab:fits1}). But after applying additionally 
the criterion item \ref{item:efsmall} this number is reduced to 22 $(V,V)$ and 25
$(V,V)_4$. However, the same argument does not remove any of the $(V,V)$
models, if $(V,V)$ is the underlying process. For a smaller number of
events, there are only a view coefficients $E,F$  smaller then
0.001 so only a smaller number of models are removed as
e.g. in Table~\ref{tab:fits2} for 10k events, 88/81 remaining $(V,V)/(V,V)_4$ models for 
the underlying $(V,S)$ process.

Since our decision making strategy depends upon the correct error estimation
of the coefficients, we have independently checked that the  
 Markov-Chain-Monte-Carlo (MCMC) method for the data fitting which yields
roughly the same errors on the coefficients.



\section{Conclusions}
\label{sec:conclusions}
In this paper we have investigated three-body decays of the
form $X\to f \bar{f} Y$ where $X$ and $Y$ are new particles and
$f$ are  SM-fermions with the aim to determine the spins of the
new particles. Here we assumed that $Y$ is a DM candidate and
escapes detection in a typical LHC detector. 

We have studied
in detail the differential width as a function of the invariant
mass of the SM-fermions for the case of heavy intermediate
particles with mass $m_I$ and expanded the width in powers of
the ratio  $\epsilon = m_X/m_I$. It turns out that general properties
such as signs or various ratios of the resulting
coefficients depend on the spin assignments of $X$ and $Y$.
From this we have developed a strategy for the spin identification
discussing various cases and testing it on large samples of
arbitrary coupling assignments. Here it turns out that one is able to
exclude several spin assignments but one does not get necessarily find
a unique solution once one has to deal with noisy data.

Although we did not find a unique solution we are 
convinced that the proposed method will be useful in practice for the
following reasons: (i) We have only investigated one particular decay
channel. However, in general several channels will be open which can
be combined. (ii) In the same spirit: we have only investigated one
decaying particle. In practice, e.g.~if supersymmetry or extra
dimensions are realized in nature, several distinct new particles 
will be produced which
eventually have to decay into the lightest of the new ones if a parity
like $R$-parity or $KK$-parity is realized.  Therefore, one will have
several different possibilities to determine the spin of the invisible
particle $Y$. (iii) Our information can be combined with other
observables, e.g.~with  cross section information. However, here one most
likely will have to assume a certain representation to which a
particular new particle belongs, e.g.~if it is
an $SU(3)$ triplet or octet.

\section*{Acknowledgements}

We thank R.~Str\"ohmer for useful discussions. This work has been supported by 
the German Ministry of Education and Research (BMBF)
under contract no. 05H09WWE. L.E. acknowledges support from the Elitenetzwerk Bayern.

\begin{appendix}
\section{Analytic Results for the Coefficients} 

The coefficients of the differential decay widths are given below. We restrict ourselves
to the case of massless SM-fermions, implying that $Z=A=0$ and, thus, the differential
widths read as
\begin{eqnarray}
\frac{d\Gamma}{d\hat{s}} =  \frac{PS}{(2\pi)^3\,\,256 \,\,m_X}  \big(  B 
+ C \hat{s} + D \hat{s}^2 + E \hat{s}^3 +F \hat{s}^4\big)
\end{eqnarray}

\subsection{Decays of new bosons} 

\label{subsec:XFFY}

The coefficients are shown with all possible diagrams and vertices in Table~\ref{tab:alltops}.
For the definition of $\epsilon$ and the various $\tau_i$ see Eq.~(\ref{eq:shortcuts}).
We give the various orders separately, e.g.
\begin{equation}
B = \sum_{j=2}^4 B_j \epsilon^j
\end{equation}
For brevity, we only explicitely write out the  higher orders for  $(S,S)$.

\paragraph{\underline{$S\rightarrow f\overline{f} S$:}\\}
\begin{flalign}
B_2  &=  128 \epsilon ^2 (\tau_Y-1)^2 (g(r,s) n(l,s)+g(l,s) n(r,s))^2 &&\nonumber\\
C_2  &=  128 \epsilon ^2 (\tau_Y-1)^2 (g(r,s) n(l,s)+g(l,s) n(r,s))^2 &&\nonumber\\
D_2  &= 0	&&
\label{eq:ss}
\end{flalign}
\begin{flalign}
B_3  &=  64 \epsilon ^3 \tau_C (\tau_Y-1)^2 c(s) (g(r,s)
   n(l,s)+g(l,s) n(r,s)) (s(l)+s(r)) &&\nonumber\\
C_3  &=  64 \epsilon ^3 \tau_C (\tau_Y-1)^2 c(s) (g(r,s)
   n(l,s)+g(l,s) n(r,s)) (s(l)+s(r)) &&\nonumber\\
D_3  &= 0	
\label{eq:ss3}
\end{flalign}
\begin{flalign}
B_4  &= \frac{16}{3} \epsilon ^4 (\tau_Y-1)^2 \left(12 g(r,s)^2 n(l,s)^2
   (\tau_Y+1)^2+12 g(l,s)^2 n(r,s)^2 (\tau_Y+1)^2
\right.
&& \nonumber\\
& \left.
+24
   g(l,s) g(r,s) n(l,s) n(r,s) (\tau_Y+1)^2+3 \tau_C^2
   c(s)^2 s(l)^2+3 \tau_C^2 c(s)^2 s(r)^2
\right.
&& \nonumber\\
& \left.
+\tau_Y^2
   c(v)^2 v(l)^2+6 \tau_Y c(v)^2 v(l)^2+c(v)^2 v(l)^2+\tau_Y^2 c(v)^2 v(r)^2
\right.
&& \nonumber\\
& \left.
+6 \tau_Y c(v)^2 v(r)^2+c(v)^2
   v(r)^2\right)   && \nonumber\\
C_4  &=  -\frac{16}{3} \epsilon ^4 (\tau_Y-1)^2 \left(-48 \tau_Y
   g(r,s)^2 n(l,s)^2-96 \tau_Y g(l,s) g(r,s) n(r,s) n(l,s)
\right.
&& \nonumber\\
& \left.
-48
   \tau_Y g(l,s)^2 n(r,s)^2-3 \tau_C^2 c(s)^2 s(l)^2-3
   \tau_C^2 c(s)^2 s(r)^2+2 \tau_Y^2 c(v)^2 v(l)^2
\right.
&& \nonumber\\
& \left.
+4
   \tau_Y c(v)^2 v(l)^2+2 c(v)^2 v(l)^2+2 \tau_Y^2
   c(v)^2 v(r)^2
\right.
&& \nonumber\\
& \left.
+4 \tau_Y c(v)^2 v(r)^2+2 c(v)^2 v(r)^2\right) &&\nonumber\\
D_4  &= -\frac{16}{3} \epsilon ^4 (\tau_Y-1)^4
   \left(-\left(v(l)^2+v(r)^2\right) c(v)^2+12 g(r,s)^2 n(l,s)^2
\right.
&& \nonumber\\
& \left.
+12
   g(l,s)^2 n(r,s)^2+24 g(l,s) g(r,s) n(l,s) n(r,s)\right)	
\label{eq:ss4}
\end{flalign}
Moreover, we get $E_j =0$ in all orders considered.

\paragraph{\underline{$S\rightarrow f\overline{f}V$:}}

\begin{flalign}
B_2  &= \frac{64}{3 \tau_Y^2} \left(g(r,s)^2 n(l,v)^2+g(l,s)^2 n(r,v)^2\right)
   \epsilon ^2 (\tau_Y-1)^2 \left(25 \tau_Y^2+6
   \tau_Y+1\right)  &&\nonumber\\
C_2  &=  \frac{128}{3 \tau_Y^2} \left(g(r,s)^2 n(l,v)^2+g(l,s)^2 n(r,v)^2\right)
   \epsilon ^2 (\tau_Y-1)^2 \left(11 \tau_Y^2-2
   \tau_Y-1\right) &&\nonumber\\
D_2  &= \frac{64}{3 \tau_Y^2} \left(g(r,s)^2 n(l,v)^2+g(l,s)^2 n(r,v)^2\right)
   \epsilon ^2 (\tau_Y-1)^4	
\label{eq:sv}
\end{flalign}

\paragraph{\underline{$V\rightarrow f\overline{f}S$:}}

\begin{flalign}
B_2  &=   \frac{64}{3} \left(g(r,v)^2 n(l,s)^2+g(l,v)^2
   n(r,s)^2\right) \epsilon ^2 (\tau_Y-1)^2 \left(\tau_Y^2+6 \tau_Y+25\right)&&\nonumber\\
C_2  &=   -\frac{128}{3} \left(g(r,v)^2 n(l,s)^2+g(l,v)^2
   n(r,s)^2\right) \epsilon ^2 (\tau_Y-1)^2 \left(\tau_Y^2+2 \tau_Y-11\right) &&\nonumber\\
D_2  &= \frac{64}{3} \left(g(r,v)^2 n(l,s)^2+g(l,v)^2
   n(r,s)^2\right) \epsilon ^2 (\tau_Y-1)^4	
\label{eq:vs}
\end{flalign}


\paragraph{\underline{$V\rightarrow f\overline{f} V$:}}
\begin{flalign}
B_2  &=  \frac{32}{3 \tau_Y^2} \epsilon ^2 (\tau_Y-1)^2 \left(g(r,v)^2 \left(3
   \tau_Y^4+16 \tau_Y^3+54 \tau_Y^2+16
   \tau_Y+3\right) n(l,v)^2
\right.
&& \nonumber\\
& \left.
-g(l,v) g(r,v) n(r,v)
   \left(3 \tau_Y^4+20 \tau_Y^3-6 \tau_Y^2+20
   \tau_Y+3\right) n(l,v)
\right.
&& \nonumber\\
& \left.
+g(l,v)^2 n(r,v)^2 \left(3
   \tau_Y^4+16 \tau_Y^3+54 \tau_Y^2+16
   \tau_Y+3\right)\right) &&\nonumber\\
C_2  &=  -\frac{32}{3 \tau_Y^2} \epsilon ^2 (\tau_Y-1)^2 \left(g(r,v)^2 \left(5
   \tau_Y^4+4 \tau_Y^3-46 \tau_Y^2+4
   \tau_Y+5\right) n(l,v)^2
\right.
&& \nonumber\\
& \left.
-g(l,v) g(r,v) n(r,v)
   \left(7 \tau_Y^4+20 \tau_Y^3+34 \tau_Y^2+20
   \tau_Y+7\right) n(l,v)
\right.
&& \nonumber\\
& \left.
+g(l,v)^2 n(r,v)^2 \left(5
   \tau_Y^4+4 \tau_Y^3-46 \tau_Y^2+4
   \tau_Y+5\right)\right) &&\nonumber\\
D_2  &= \frac{32}{3 \tau_Y^2} \epsilon ^2 (\tau_Y-1)^4 \left(g(r,v)^2
   \left(\tau_Y^2-6 \tau_Y+1\right)
   n(l,v)^2
\right.
&& \nonumber\\
& \left.
-g(l,v) g(r,v) n(r,v) \left(5 \tau_Y^2+6 \tau_Y+5\right) n(l,v)
\right.
&& \nonumber\\
& \left.
+g(l,v)^2 n(r,v)^2
   \left(\tau_Y^2-6 \tau_Y+1\right)\right)	&&\nonumber\\
E_2  &=  \frac{32}{3 \tau_Y^2} \left(g(r,v)^2 n(l,v)^2+g(l,v) g(r,v) n(r,v)
   n(l,v)+g(l,v)^2 n(r,v)^2\right) \epsilon ^2 (\tau_Y-1)^6 
\label{eq:vv}
\end{flalign}

\subsection{Decays of new fermions}
\label{subsec:mffm}
As noted before, in this case only the 4th order in $\epsilon$ contributes.
We split the various coefficients according to the different topologies
considered, e.g. the scalar contributions to $B_4$ are
\begin{equation}
B_4 = B_s + B_s' + B_{s,m} 
\end{equation}
where
\begin{eqnarray*}
B_s:& \mbox{top. 1+2 with intermediate scalars}&\\
B_s':& \mbox{top. 3 with intermediate scalars}&\\
B_{s,m}:& \mbox{interference term of top. (1+2)+3 with intermediate scalars}
\end{eqnarray*}
For intermediate vector bosons the index $v$ is used. For the
interference terms between scalars and vector bosons the index $(sv)$
is used in case of topologies 1+2, $(sv1)$ for the scalars of topology 1+2 and
vector bosons of topology 3, $(s1v)$ for the scalars of topology 3 and
vector bosons of topology 1+2. 
Moreover we find that the interference vanishes
if both, scalars and vector bosons stem from the third topology because we 
have $m_f=0$ for the SM-fermions.

\paragraph{\underline{Intermediate scalars:}}

\begin{flalign}
B_s &= \frac{64}{3}  (\tau_Y-1)^2 \left(2
   \left((\tau_Y (\tau_Y+6)+1)
   n(l,s)^2+(\tau_Y (\tau_Y+3)+1)
   n(r,s)^2\right) g(l,s)^2
\right.&&\nonumber\\
& \left.
-(\tau_Y
   (\tau_Y+6)+1)g(r,s) n(l,s)
   n(r,s) g(l,s)+2g(r,s)^2
   \left((\tau_Y (\tau_Y+3)+1)
   n(l,s)^2
\right.\right.&&\nonumber\\
& \left.\left.
+(\tau_Y (\tau_Y+6)+1)
   n(r,s)^2\right)\right)  &&\nonumber\\
C_s &= -\frac{64}{3}  (\tau_Y-1)^2 \left(-2
   g(l,s)g(r,s) n(l,s) n(r,s)
   (\tau_Y+1)^2
\right.&&\nonumber\\
& \left.
+g(r,s)^2 \left(((\tau_Y-4) \tau_Y+1) n(l,s)^2+((\tau_Y-10)
   \tau_Y+1) n(r,s)^2\right)
\right.&&\nonumber\\
& \left.
+g(l,s)^2
   \left(((\tau_Y-10) \tau_Y+1)
   n(l,s)^2+((\tau_Y-4) \tau_Y+1)
   n(r,s)^2\right)\right) &&\nonumber\\
D_s &= -\frac{64}{3}  (\tau_Y-1)^4
   \left(\left(n(l,s)^2+n(r,s)^2\right)
   g(l,s)^2
\right.&&\nonumber\\
& \left.
+g(r,s) n(l,s) n(r,s)
   g(l,s)+g(r,s)^2
   \left(n(l,s)^2+n(r,s)^2\right)\right)
\end{flalign}
\begin{flalign}
B_s' &= 32  (\tau_Y-1)^2 \left((\tau_Y+1)^2
   d(l,s)^2
\right.&&\nonumber\\
& \left.
+8 \tau_Y d(r,s)
   d(l,s)+(\tau_Y+1)^2 d(r,s)^2\right)
   \left(s(l)^2+s(r)^2\right) &&\nonumber\\
C_s' &= 128  (\tau_Y-1)^2 \tau_Y
   (d(l,s)+d(r,s))^2
   \left(s(l)^2+s(r)^2\right) &&\nonumber\\
D_s' &= -32  (\tau_Y-1)^4
   \left(d(l,s)^2+d(r,s)^2\right)
   \left(s(l)^2+s(r)^2\right) 
\end{flalign}
\begin{flalign}
B_{s,m} &= -32  (\tau_Y-1)^2 (g(r,s)
   n(l,s)+g(l,s) n(r,s))
   \left(d(r,s) \left(s(l) (\tau_Y+1)^2+4
   \tau_Y s(r)\right)
\right.&&\nonumber\\
& \left.
+d(l,s)
   \left(s(r) (\tau_Y+1)^2+4 \tau_Y
   s(l)\right)\right) &&\nonumber\\
C_{s,m} &= -128  (\tau_Y-1)^2 \tau_Y
   (g(r,s) n(l,s)+g(l,s) n(r,s))
   (d(l,s)+d(r,s)) (s(l)+s(r)) &&\nonumber\\
D_{s,m} &= 32  (\tau_Y-1)^4 (g(r,s)
   n(l,s)+g(l,s) n(r,s)) (d(r,s)
   s(l)+d(l,s) s(r)) &&
\end{flalign}

\paragraph{\underline{Intermediate vector bosons:}}

\begin{flalign}
B_v &= \frac{256}{3}  (\tau_Y-1)^2 \left(6
   g(l,v) g(r,v) n(l,v) n(r,v)
   (\tau_Y+1)^2
\right.&&\nonumber\\
& \left.
+g(l,v)^2 \left(2 (\tau_Y
   (\tau_Y+6)+1) n(l,v)^2+3 (\tau_Y+1)^2
   n(r,v)^2\right)
\right.&&\nonumber\\
& \left.
+g(r,v)^2 \left(3 (\tau_Y+1)^2 n(l,v)^2+2 (\tau_Y (\tau_Y+6)+1) n(r,v)^2\right)\right) &&\nonumber\\
C_v &=  \frac{256}{3}  (\tau_Y-1)^2 \left(\left(12
   \tau_Y n(r,v)^2-((\tau_Y-10) \tau_Y+1) n(l,v)^2\right) g(l,v)^2
\right.&&\nonumber\\
& \left.
+24 \tau_Y g(r,v) n(l,v) n(r,v)
   g(l,v)+g(r,v)^2 \left(12 \tau_Y
   n(l,v)^2-((\tau_Y-10) \tau_Y+1)
   n(r,v)^2\right)\right) &&\nonumber\\
D_v &= -\frac{256}{3}  (\tau_Y-1)^4
   \left(\left(n(l,v)^2+3 n(r,v)^2\right)
   g(l,v)^2
\right.&&\nonumber\\
& \left.
+6 g(r,v) n(l,v)
   n(r,v) g(l,v)+g(r,v)^2 \left(3
   n(l,v)^2+n(r,v)^2\right)\right) &&
\end{flalign}
\begin{flalign}
B_v' &= \frac{256}{3}  (\tau_Y-1)^2 \left((\tau_Y (\tau_Y+3)+1) d(l,v)^2-6 \tau_Y
   d(r,v) d(l,v)
\right.&&\nonumber\\
& \left.
+(\tau_Y (\tau_Y+3)+1) d(r,v)^2\right)
   \left(v(l)^2+v(r)^2\right) &&\nonumber\\
C_v' &= -\frac{128}{3}  (\tau_Y-1)^2 \left(((\tau_Y-4) \tau_Y+1) d(l,v)^2
\right.&&\nonumber\\
&\left.
+12 \tau_Y
   d(r,v) d(l,v)+((\tau_Y-4) \tau_Y+1) d(r,v)^2\right)
   \left(v(l)^2+v(r)^2\right) &&\nonumber\\
D_v' &= -\frac{128}{3}  (\tau_Y-1)^4
   \left(d(l,v)^2+d(r,v)^2\right)
   \left(v(l)^2+v(r)^2\right) &&
\end{flalign}
\begin{flalign}
B_{v,m} &= \frac{512}{3}  (\tau_Y-1)^2 (\tau_Y
   (\tau_Y+6)+1) (d(l,v)-d(r,v))
   &&\nonumber\\
&(g(l,v) n(l,v) v(l)-g(r,v)
   n(r,v) v(r)) &&\nonumber\\
C_{v,m} &= -\frac{256}{3}  (\tau_Y-1)^2 ((\tau_Y-10) \tau_Y+1) (d(l,v)-d(r,v))
   &&\nonumber\\
&(g(l,v) n(l,v) v(l)-g(r,v)
   n(r,v) v(r)) &&\nonumber\\
D_{v,m} &= -\frac{256}{3}  (\tau_Y-1)^4
   (d(l,v)-d(r,v)) (g(l,v) n(l,v)
   v(l)-g(r,v) n(r,v) v(r)) &&
\end{flalign}

\paragraph{\underline{Interference terms between scalars and vector bosons:}}

\begin{flalign}
B_{sv} &= -\frac{512}{3}  (\tau_Y-1)^2 (n(l,v) (3
   \tau_Y g(l,s) g(r,v)
   n(r,s)+g(r,s) (3 \tau_Y
   g(r,v) n(l,s)
&&\nonumber\\
& 
+(\tau_Y (\tau_Y+6)+1) g(l,v) n(r,s)))+((\tau_Y
   (\tau_Y+6)+1) g(l,s) g(r,v)
   n(l,s)
&&\nonumber\\
&
+3 \tau_Y g(l,v)
   (g(r,s) n(l,s)+g(l,s)
   n(r,s))) n(r,v)) &&\nonumber\\
C_{sv} &= \frac{256}{3}  (\tau_Y-1)^2 (n(l,v)
   (g(r,s) (((\tau_Y-10) \tau_Y+1)
   g(l,v) n(r,s)
&&\nonumber\\
&
-6 \tau_Y
   g(r,v) n(l,s))-6 \tau_Y
   g(l,s) g(r,v) n(r,s))+(((\tau_Y-10) \tau_Y+1) g(l,s) g(r,v)
   n(l,s)
&&\nonumber\\
&-6 \tau_Y g(l,v)
   (g(r,s) n(l,s)+g(l,s)
   n(r,s))) n(r,v)) &&\nonumber\\
D_{sv} &= \frac{256}{3}  (\tau_Y-1)^4 (g(l,v)
  g(r,s) n(l,v) n(r,s)+g(l,s)
   g(r,v) n(l,s) n(r,v)) &&
\end{flalign}
\begin{flalign}
B_{sv1} &= -\frac{256}{3}  (\tau_Y-1)^2 (\tau_Y
   (\tau_Y+6)+1) (d(l,v)-d(r,v))
&&\nonumber\\
&   (g(r,s) n(r,s) v(l)-g(l,s)
   n(l,s) v(r)) &&\nonumber\\
C_{sv1} &= \frac{128}{3}  (\tau_Y-1)^2 ((\tau_Y-10)
   \tau_Y+1) (d(l,v)-d(r,v))
&&\nonumber\\
&  (g(r,s) n(r,s) v(l)-g(l,s)
   n(l,s) v(r)) &&\nonumber\\
D_{sv1} &= \frac{128}{3}  (\tau_Y-1)^4
   (d(l,v)-d(r,v)) (g(r,s) n(r,s)
   v(l)-g(l,s) n(l,s) v(r)) &&
\end{flalign}
\begin{flalign}
B_{s1v} &=  128  (\tau_Y-1)^2 (g(r,v)
   n(l,v)+g(l,v) n(r,v))
   \left(d(l,s) \left(s(l) (\tau_Y+1)^2+4
   \tau_Y s(r)\right)
\right.
&&\nonumber\\
& \left.
+d(r,s)
   \left(s(r) (\tau_Y+1)^2+4 \tau_Y
   s(l)\right)\right)&&\nonumber\\
C_{s1v} &= 512  (\tau_Y-1)^2 \tau_Y
   (g(r,v) n(l,v)+g(l,v) n(r,v))
&&\nonumber\\
& 
 (d(l,s)+d(r,s)) (s(l)+s(r)) &&\nonumber\\
D_{s1v} &= -128  (\tau_Y-1)^4 (g(r,v)
   n(l,v)+g(l,v) n(r,v)) (d(l,s)
   s(l)+d(r,s) s(r)) &&
\end{flalign}

\end{appendix}

\end{document}